\documentclass[manuscript, 12pt]{elsarticle}
\usepackage{hyperref}

\usepackage{graphics}
\usepackage{amssymb}
 
 \usepackage{amsmath}
 \usepackage{textcomp}
 \usepackage{gensymb}
 
 \usepackage{color}

\usepackage{lineno}

\usepackage{TIGcommands}

\journal{Journal of Computational Physics}

\begin{document}

\begin{frontmatter}



\title{Gauss's Law Satisfying Energy-Conserving Semi-Implicit Particle-in-Cell Method}


\author[label1]{Yuxi Chen\footnote{Corresponding author. Email address: yuxichen@umich.edu}}
\author[label1]{G\'abor T\'oth}

\address[label1]{
Center for Space Environment Modeling, University of Michigan,
Ann Arbor, Michigan 48109, USA}

\begin{abstract}
The Energy Conserving Semi-Implicit Method (ECSIM) introduced by Lapenta (2017) has many advantageous properties compared to the classical semi‐implicit and explicit PIC methods. Most importantly, energy conservation eliminates the growth of the finite grid instability. We have implemented ECSIM in a different and more efficient manner than the original approach. More importantly, we have addressed two major shortcomings of the original ECSIM algorithm: there is no mechanism to enforce Gauss's law and there is no mechanism to reduce the numerical oscillations of the electric field. A classical approach to satisfy Gauss's law is to modify the electric field and its divergence using either an elliptic or a parabolic/hyperbolic correction based on the Generalized Lagrange Multiplier method. This correction, however, violates the energy conservation property, and the oscillations related to the finite grid instability reappear in the modified ECSIM scheme. We invented a new alternative approach: the particle positions are modified instead of the electric field in the correction step. Displacing the particles slightly does not change the energy conservation property, while it can satisfy Gauss's law by changing the charge density. We found that the new Gauss's Law satisfying Energy Conserving Semi-Implicit Method (GL-ECSIM) produces superior results compared to the original ECSIM algorithm. In some simulations, however, there are still some numerical oscillations present in the electric field. We attribute this to the simple finite difference discretization of the energy conserving implicit electric field solver. We modified the spatial discretization of the field solver to reduce these oscillations while only slightly violating the energy conservation properties. We demonstrate the improved quality of the GL-ECSIM method with several tests.

\end{abstract}


\end{frontmatter}

\section{Introduction}

Conservation properties play an important role to avoid numerical
instabilities for the particle-in-cell (PIC) method. The explicit PIC 
method, which is widely used due to its simplicity, conserves the total
 moment but tends to increase the total energy of the system by numerical heating. 
 The implicit PIC method, which relaxes the temporal and spatial stability constraints,
tends to decrease the system energy by numerical cooling. Fully implicit PIC schemes
can achieve energy conservation by solving for 
the particle motions and electro-magnetic fields at the same time via a non-linear 
Newton-Krylov iterative solver \citep{Markidis:2011,Chen:2011,Chen:2015,Chacon:2016}. 
Recently, Lapenta \cite{Lapenta:2017a} proposed an Energy Conserving 
Semi-Implicit Method (ECSIM) that conserves energy by ensuring the current used for
electric field updating is the same as the current produced by moving particles.

Another important conservation law is related to Gauss's law:
\begin{linenomath}\begin{equation} \label{eq:Gauss}
\nabla \cdot \mathbf{E} = 4\pi\rho
\end{equation} \end{linenomath}
where $\mathbf{E}$ is the electric field and $\rho$ is the electric charge density. Analytically, Gauss's law will be satisfied if the initial condition satisfies it and Amp\`ere's law and the charge conservation equations are solved exactly. Amp\`ere's law describes the evolution equation for the electric field: 
\begin{linenomath}\begin{equation} \label{eq:Ampere}
\frac{\partial \mathbf{E}}{\partial t} = c\nabla \times \mathbf{B}
- 4\pi\mathbf{J} 
\end{equation} \end{linenomath}
where $\mathbf{J}$ is the current density, $\mathbf{B}$ is the magnetic field vector and $c$ is the speed of light. The charge density evolves according to
\begin{linenomath}\begin{equation} \label{eq:charge-conservation}
\frac{\partial \rho}{\partial t} + \nabla \cdot \mathbf{J} = 0
\end{equation} \end{linenomath}
Taking the divergence of Amp\`ere's law and using the charge conservation leads to
\begin{linenomath}\begin{equation} \label{eq:Gauss-time}
\frac{\partial \nabla \cdot \mathbf{E}}{\partial t} 
= 4\pi\frac{\partial \rho}{\partial t}
\end{equation} \end{linenomath}
which means that Gauss's law is maintained as long as it holds initially.

The electromagnetic PIC methods usually update the electric field by
solving Amp\`ere’s law from the magnetic field and the current on a grid. This current is interpolated to the grid from the
particles and does not necessarily satisfy the charge conservation equation. This discrepancy
may accumulate and lead to significant violation of Gauss's law. Two classes of methods have been
proposed to solve this numerical issue. One approach is enforcing the electric field to satisfy
Gauss's law by applying a correction term to the electric field equation. 
The correction can be applied as an extra correction step, or added to the electric field solver directly. 
Boris'  popular $\nabla \cdot \mathbf{E}$ error correction method \cite{Boris:1970,Birdsall:2014}solves
a Poisson equation and reduces the error in Gauss's law to the iteration tolerance level.
Marder \cite{Marder:1987} and Langdon \cite{Langdon:1992} reduce the computational cost by
replacing the Poisson solver with a local fix. Marder \cite{Marder:1987} calls the correction term
as `pseudo-current'. The idea of electric field correction is generalized by Assous et al. \cite{Assous:1993}
and Munz et al. \cite{Munz:2000} in a generalized Lagrange multiplier (GLM) numerical framework, where
new variables are introduced to the Maxwell's equations to constrain the errors in Gauss's law.
The other class of methods does not require any electric field correction. Instead, these methods
carefully design the algorithm so that the current assigned to the electric field solver satisfies
the charge conservation equation and hence Gauss's law automatically .
Buneman \cite{Buneman:1968} developed the `zero-order current weighting' algorithm, which
uses an impulse current assignment when a particle crosses a cell boundary. 
Similarly, Morse and Nielson \cite{Morse:1971} proposed the `first-order current weighting'
method, where the current is assigned by area weighting and the particle motion
is divided into two or three orthogonal moves. 
Villasenor and Buneman \cite{Villasenor:1992} introduced another area weighting method which does 
not require the orthogonal motion splitting. This scheme is generalized to any form-factor
by Esirkepov \cite{Esirkepov:2001}. Umeda et al. \cite{Umeda:2003} developed 
an algorithm similar to Villasenor and Buneman \cite{Villasenor:1992} but assumes
the particle trajectory is zigzag. Sokolov \cite{Sokolov:2013altor} introduced a
method to conserve charge using an alternating order form-factor. 
Eastwood \cite{Eastwood:1991,Eastwood:1995} presented a general description of the charge 
conserving scheme for Cartesian and curvilinear grids. Besides these two classes of techniques, 
Chen and Chac\'on \cite{Chen:2011, Chen:2015, Chacon:2016} designed a class of fully implicit 
methods that conserve charge and energy at the same time. 

The Energy Conserving Semi-Implicit Method (ECSIM) \cite{Lapenta:2017a} conserves the energy up 
to the iteration tolerance. It is faster than the explicit PIC methods due to the relaxed 
temporal and spatial resolution constraints. ECSIM is also more efficient than the fully explicit methods, 
because ECSIM does not require the particles to be involved during the iterations.
Lapenta\cite{Lapenta:2017a} demonstrated that ECSIM is about one order faster than a fully
implicit PIC code for 1D problems when the same grid resolution and number of 
particles are used (Table 1 and Table 2 of \cite{Lapenta:2017a}).
A potential flaw of ECSIM is the lack of any mechanism ensuring the satisfaction of Gauss's law. 
The violation of Gauss's law may generate numerical artifacts. The electric field
correction method can be easily applied to ECSIM to improve the charge conservation,
but it destroys the energy conservation property, and more importantly it does not behave well 
for certain cases as we will demonstrate in this paper. It is also not trivial 
to design a current assignment algorithm to satisfy both energy conservation and charge 
conservation at the same time for the semi-implicit moment method. 

We have successfully applied the semi-implicit PIC algorithm implemented into the iPIC3D code \cite{Markidis:2010} 
to large-scale kinetic simulations in recent years \cite{Daldorff:2014,Toth:2016,Chen:2017,Toth:2017}. 
We found that the code may create artificial oscillations in the electric field and heat the particles 
numerically, which needs to be alleviated by smoothing the electric field \cite{Chen:2017,Toth:2017,Gonzalez:2018}.
Smoothing will, of course, make the solution more diffusive. ECSIM provides another option to eliminate the numerical heating by enforcing conservation of energy.
We implemented the ECSIM algorithm into iPIC3D in an efficient way, which is described in section \ref{section:method},  
but we found that ECSIM may create other numerical issues related to the violation of Gauss's law. In this paper, we introduce the novel idea to correct the particle locations at the end of each
computational cycle to satisfy Gauss's law for the ECSIM algorithm. The correction keeps 
the energy conservation property of ECSIM because it changes neither the kinetic
energy of each particle nor the electromagnetic field energy. Since there are usually at least
dozens of macro-particles per cell, the displacement of each particle required to eliminate
the errors in Gauss's law is not unique. In order to minimize the displacements, we apply a generalized Lagrange
multiplier to minimize the total displacements of the macro-particles while satisfying Gauss's law at every grid cell.  This
correction is accurate but also computationally intensive. To reduce the computational cost, 
we also designed another two alternative approximate correction methods, which do not eliminate 
the error entirely, but can suppress the growth of the error effectively and are computationally less expensive.
The three variants of this novel Gauss's Law satisfying Energy-Conserving Semi-Implicit Method (GL-ECSIM) 
are described in section \ref{section:method}.  

 We note that even though this particle position correction method is designed to 
 improve the performance of ECSIM, the same idea can be easily applied to any other PIC algorithm. Correcting the particle positions instead of the electric field may be advantageous, because in general the field quantities are smoother and have less error than the particle related quantities, like charge density. Correcting the particle positions is likely to remove actual errors (compared to an exact solution), while correcting the electric field may push the errors in the particle positions into the electric field. 
 
Besides the Gauss's law satisfaction issue, we also found 
ECSIM may produce short-wavelength  oscillation due to the simple spatial discretization used for the 
electric field solver.  Section \ref{section:method} also discusses the modifications that are 
necessary to suppress the oscillations. Numerical tests in section \ref{section:test} justify
the necessity of improving the charge conservation property and other modifications, and 
demonstrate the quality of our algorithm. Finally, section \ref{section:conclusion} presents the conclusions.

\section{The Gauss's law satisfying energy-conserving semi-implicit method (GL-ECSIM)}
\label{section:method}

\subsection{The electric field solver}
GL-ECSIM is based on the Energy-Conserving Semi-Implicit Method (ECSIM)
developed by Lapenta \cite{Lapenta:2017a}. ECSIM uses a staggered grid, where the
electric field is defined at cell nodes, and the magnetic field is stored at cell centers. 
The position and velocity of a macro-particle are staggered in time, i.e., the particle velocity is at the integer
time stage and the location is at the half time stage. Lapenta \cite{Lapenta:2017a} updates the 
electric field and magnetic field at the same time by an implicit solver:

\begin{linenomath} \begin{eqnarray}
\frac{\mathbf{B}^{n+1}-\mathbf{B}^{n}}{\Delta t} &=&
      -c\nabla \times \mathbf{E}^{n+\theta} 
\label{eq:theta-B}\\
\frac{\mathbf{E}^{n+1}-\mathbf{E}^{n}}{\Delta t} &=& 
c\nabla \times \mathbf{B}^{n+\theta}  - 4\pi \mathbf{\bar{J}}
\label{eq:theta-E}
\end{eqnarray}\end{linenomath}
where $\mathbf{\bar{J}}$ is the predicted current at $n+\frac{1}{2}$ time stage, and it
depends on the unknown electric field $\mathbf{E}^{n+\theta}$. The definition 
of current $\mathbf{\bar{J}}$ can be found in \cite{Lapenta:2017a}.
The value at time level $n+\theta$ is defined as a linear combination of the
 values at the $n$ and $n+1$ stages such that:
\begin{linenomath} \begin{eqnarray}
\mathbf{E}^{n+\theta} = (1-\theta)\mathbf{E}^{n} + \theta\mathbf{E}^{n+1} \label{eq:theta-E1}\\
\mathbf{B}^{n+\theta} = (1-\theta)\mathbf{B}^{n} + \theta\mathbf{B}^{n+1}
\label{eq:theta-B1}
\end{eqnarray}\end{linenomath}
Instead of solving for $\mathbf{E^{n+1}}$ and $\mathbf{B}^{n+1}$ at the same time, we replace
 $\mathbf{B}^{n+1}$ and $\mathbf{E}^{n+1}$ in eq.(\ref{eq:theta-B}) and eq.(\ref{eq:theta-E})
  with linear combinations of  $\mathbf{B}^{n}$, $\mathbf{B}^{n+\theta}$ and  $\mathbf{E}^{n}$, $\mathbf{E}^{n+\theta}$, 
 respectively, express $\mathbf{B}^{n+\theta}$ from
eq.(\ref{eq:theta-B}) and substitute this into eq.(\ref{eq:theta-E}) to obtain an equation that only contains the electric field as unknowns:
\begin{linenomath}\begin{eqnarray}
\mathbf{E}^{n+\theta} + \delta^2 \left[ \nabla(\nabla \cdot \mathbf{E}^{n+\theta}) - 
\nabla ^2 \mathbf{E}^{n+\theta}\right] =\mathbf{E}^{n} 
   + \delta\left(\nabla \times \mathbf{B}^n - \frac{4\pi}{c}\mathbf{\bar{J}}\right),
\label{eq:ECSIM-E}
\end{eqnarray}\end{linenomath}
where $\delta = c \theta \Delta t$, and the identity
$\nabla\times\nabla\times\mathbf{E} = \nabla(\nabla \cdot \mathbf{E}) - \nabla ^2 \mathbf{E}$ is used, which also holds numerically for the specific spatial discretization of the ECSIM algorithm.
After $\mathbf{E}^{n+\theta}$ is obtained, the magnetic field at time level $n+1$ can be easily calculated from eq.(\ref{eq:theta-B}).
Solving eq.(\ref{eq:ECSIM-E}) is equivalent to solving eqs.(\ref{eq:theta-B}) $-$ (\ref{eq:theta-E}) analytically.
 But there are some numerical advantages of solving eq.(\ref{eq:ECSIM-E}) 
instead of eq.(\ref{eq:theta-B}) - eq.(\ref{eq:theta-E}):
\begin{itemize}
\item The number of unknown variables per grid cell is reduced from 6 to 3.
\item Eq.(\ref{eq:ECSIM-E}) transfers two curl operators in eqs.(\ref{eq:theta-E}) $-$ (\ref{eq:theta-B})
into a Laplacian and a gradient-divergence term. The Laplacian operator is diagonally 
dominant and helps to speed up the convergence.
This transformation is proposed by Chac\'on and Knoll \cite{Chacon:2003}, and known as the 
'physics-based' preconditioner.
\end{itemize}
We use the GMRES iterative scheme to solve eq.(\ref{eq:ECSIM-E}). The magnetic field is updated from eq.(\ref{eq:theta-B}) after the electric field is obtained.

As it has been pointed out by Lapenta \cite{Lapenta:2017a}, the exact energy conservation 
can be achieved only if $\theta=0.5$ and proper spatial discretizations are used. 
But simulations with $\theta=0.5$ have more noise than the simulations with $\theta=1$ \cite{Lapenta:2017a}.
Our tests in section \ref{section:test} confirm that simulations with $\theta=0.5$ may create 
numerical waves. We propose using $\theta=0.51$ instead. This choice sacrifices the energy 
conservation a little bit, but improves the robustness significantly. Our observations are 
consistent with Tanaka's work \cite{Tanaka:1988,Tanaka:1995b} that pointed out that 
$\theta > 0.5$ damps the light waves and the Langmuir oscillations in a semi-implicit PIC method that uses a temporal discretization similar to ECSIM. 

\subsection{The pseudo-current}
The ECSIM method is the further development of the iPIC3D code \cite{Markidis:2010}, which also solves an electric 
field equation similar to eq.(\ref{eq:ECSIM-E}). Our numerical tests show iPIC3D satisfies Gauss's law better than the
ECSIM method in general, because iPIC3D incorporates a `pseudo-current' \cite{Marder:1987} term into its electric field solver. 
To illustrate this point, we add a $(c\Delta t)^2\nabla \nabla \cdot{\mathbf{E}^{n+1}}$ term to
both sides of eq.(15) in
\cite{Markidis:2010}, which is the electric field equation iPIC3D solves, and re-organize it:
 \begin{linenomath}\begin{equation}
\begin{split}
\mathbf{E}^{n+1} + (c\Delta t)^2 \left[ \nabla(\nabla \cdot \mathbf{E}^{n+1}) - \nabla ^2 \mathbf{E}^{n+1}\right] = & 
 \mathbf{E}^{n} + c\Delta t(\nabla \times \mathbf{B}^n - \frac{4\pi}{c}\mathbf{\bar{J}}) 
 \\ & -(c\Delta t)^2 \nabla(4\pi \rho^{n+1} - \nabla \cdot \mathbf{E}^{n+1}) ,
\end{split}
\label{eq:iPIC3D-E}
\end{equation}\end{linenomath}
where $\bar{\mathbf{J}}$ is the current at half time stage, just as the current 
in eq.(\ref{eq:ECSIM-E}) but it is calculated in a different way,
 and $\rho ^{n+1}$ is the estimated net charge density at the $n+1$ stage:
\begin{linenomath}
\begin{eqnarray}
\mathbf{\bar{J}} = \mathbf{\hat{J}} + \frac{\chi ^n}{4\pi \Delta t} \mathbf{E}^{n+1},\\
\rho ^{n+1} = \rho ^{n} - \Delta t \nabla \cdot \bar{\mathbf{J}}.\label{eq:continuity}
\end{eqnarray}
\end{linenomath}
The definition of $\hat{\mathbf{J}}$ and $\chi^n$ can be found in \cite{Markidis:2010}.
The last two terms in eq.(\ref{eq:iPIC3D-E}), which are the difference between the charge 
and the divergence of the electric field, correspond to the 'pseudo-current' and diffuse the 
errors away. The diffusion effect can be seen by taking the divergence of the 
semi-discretized equation eq.(\ref{eq:iPIC3D-E}), and applying the equality 
$\nabla\times\nabla\times\mathbf{E} = \nabla(\nabla \cdot \mathbf{E}) - \nabla ^2 \mathbf{E}$ 
and the electric charge continuity equation eq.(\ref{eq:continuity}):
\begin{linenomath}
\begin{eqnarray}
\frac{(\nabla \cdot \mathbf{E}^{n+1} - 4\pi \rho^{n+1}) - (\nabla \cdot \mathbf{E}^{n} - 4\pi \rho^{n}) }{\Delta t} = c^2 \Delta t \nabla ^2(\nabla \cdot \mathbf{E}^{n+1} - 4\pi \rho^{n+1}),
\label{eq:diffusion}
\end{eqnarray}
\end{linenomath}
which is a diffusion equation for the error in Gauss's law. A more detailed analysis can be found in Marder \cite{Marder:1987}.

When $\theta=1$ is chosen for the ECSIM solver eq.(\ref{eq:ECSIM-E}), it is very similar to the
iPIC3D solver eq.(\ref{eq:iPIC3D-E}) except that there is a pseudo-current term in the iPIC3D
solver and these two PIC methods use different algorithms to calculate the current $\bar{\mathbf{J}}$.
The pseudo-current method can be applied to the ECSIM's electric field solver eq.(\ref{eq:ECSIM-E})
as well. 
 We add the term  $-\delta^2 \nabla(4\pi \rho^{n+1/2} - \nabla \cdot \mathbf{E}^{n+\theta})$ 
 to the right-hand
 side of eq.(\ref{eq:ECSIM-E}) and move the $\nabla \cdot \mathbf{E}^{n+\theta}$ term to the left-hand side to obtain:
 \begin{linenomath}
 \begin{eqnarray}
\mathbf{E}^{n+\theta} + \delta^2 \left[(1-c_{pc}) \nabla(\nabla \cdot \mathbf{E}^{n+\theta}) 
 - \nabla ^2 \mathbf{E}^{n+\theta}\right] &=&
\mathbf{E}^{n} + \delta\left(\nabla \times \mathbf{B}^n - \frac{4\pi}{c}\bar{\mathbf{J}}\right)  \nonumber\\
&&-c_{pc}\delta^2\nabla(4\pi \rho ^{n+\frac{1}{2}})
\label{eq:E-pse}
\end{eqnarray}
\end{linenomath}
where $c_{pc}$ is the coefficient of the pseudo-current. It is easy to implement this pseudo-current
term, because the field $\mathbf{E}^{n+\theta}$ is already part of the field solver and the net charge 
$\rho ^{n+\frac{1}{2}}$ can be calculated from the particles in advance. We use $\mathbf{E}^{n+\theta}$ and 
$\rho ^{n+\frac{1}{2}}$ to form the pseudo-current term for simplicity. $\mathbf{E}^{n+\theta}$ and 
$\rho ^{n+\frac{1}{2}}$ are not necessarily at the same time stage unless $\theta = 0.5$. 
In section \ref{section:test}, we show that the pseudo-current scheme does not work well for 
the ECSIM method in general, because it ruins the energy conservation.

\subsection{Particle position correction}
\label{section:particle correction}

The electric field correction methods, such as the 'pseudo-current' method, modify the electric field 
to reduce the discrepancy in Gauss's law. If most of the error in Gauss's law is due to 
the inaccuracy of the net charge, which comes from the particle mover, the field correction method will not work well even though Gauss's law is satisfied formally. 

In this section, we introduce a new idea of displacing the particles to satisfy Gauss's law. The displacement
is done at the end of each computational cycle after each particle has updated its velocity and position. 
Since neither the electromagnetic field nor the particle velocity are changed by the particle position
correction, the energy conservation still holds. The particle position correction method 
can be accurate or approximate. The accurate correction need to calculate the particle displacement 
carefully to perfectly satisfy Gauss's law at every grid cell, 
while the approximate correction just moves 
the particles in the right direction to reduce the error in Gauss's law.

\subsubsection{The accurate correction}
\begin{figure}
\begin{center}
\includegraphics[width=\textwidth]{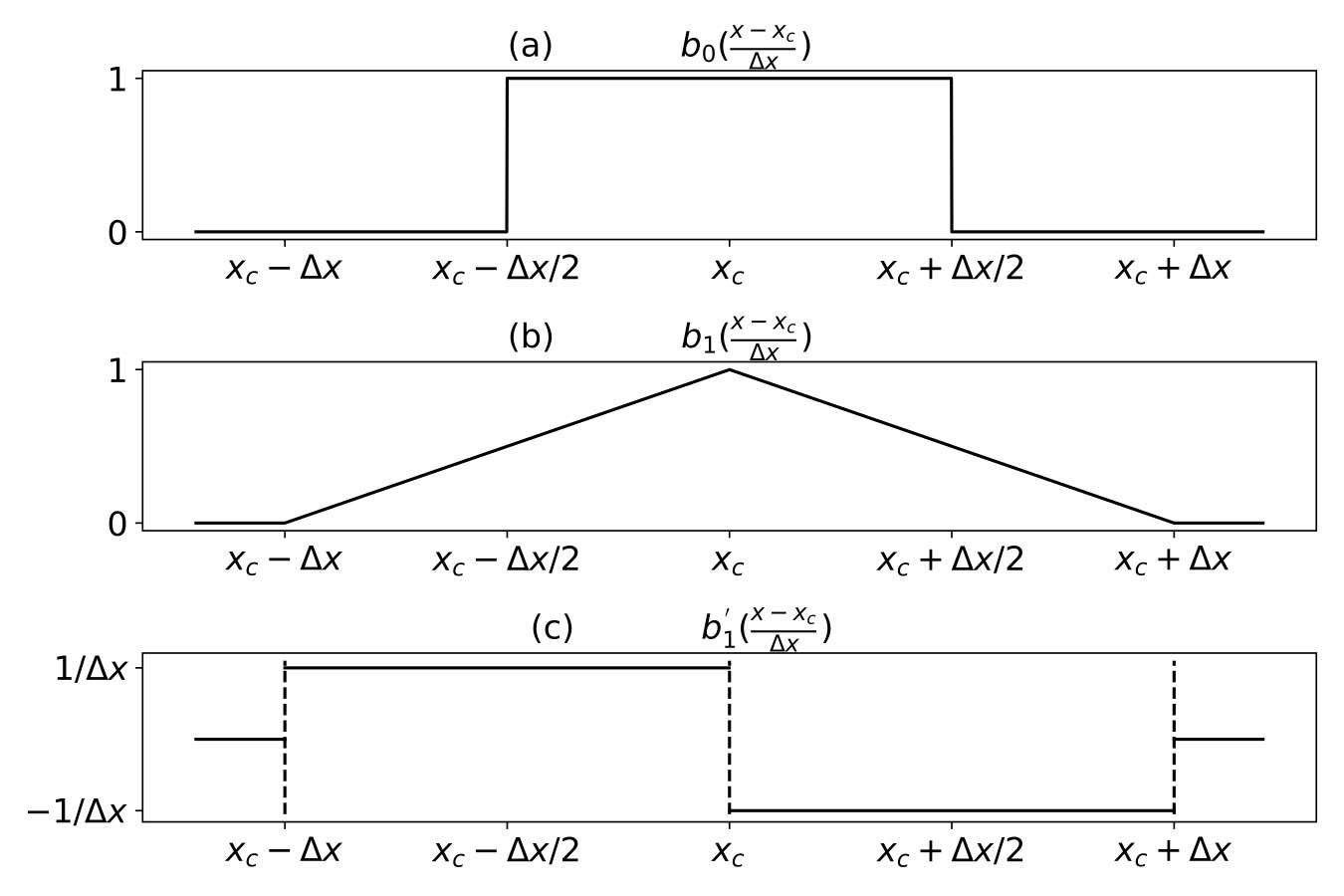}
\caption{The B-spline functions and the derivative of $b_1$. The $b_0$ spline at the top is used in the shape function $S$ while the $b_1$ spline in the middle is used for the interpolation function $W$. The derivative of $b_1$ at the bottom is needed in the gradient of $W$.
\label{fig:b-splines}
}
\end{center}
\end{figure}

In one computational cycle, the electromagnetic field is updated from $\mathbf{E}_g^n$ and $\mathbf{B}_c^n$
to $\mathbf{E}_g^{n+1}$ and $\mathbf{B}_c^{n+1}$, the particle's velocity is updated from $\mathbf{v}_p^n$
to $\mathbf{v}_p^{n+1}$ and the particle is moved from $\mathbf{x}_p^{n+\frac{1}{2}}$ to 
$\tilde{\mathbf{x}}_p^{n+\frac{3}{2}}$. We use subscripts $p$, $c$ and $g$ to represent particles,
cell centers and cell nodes, respectively. The tilde marks the values before the correction.

We use the node electric field and cell center net charge to evaluate the error in Gauss's law. 
The net charge density at the cell center is interpolated from particles. For example, 
\begin{linenomath}
 \begin{equation}
\rho_c^{n+\frac{1}{2}} = \sum_p q_pW(\mathbf{x}_p^{n+\frac{1}{2}}-\mathbf{x}_c)
\label{eq:rhoc}
\end{equation}
\end{linenomath}
where $\rho_c^{n+\frac{1}{2}}$ is the cell center net charge density at the $n+\frac{1}{2}$ time stage, 
$q_p$ is the charge of a macro-particle and $W(\mathbf{x}_p^{n+\frac{1}{2}}-\mathbf{x}_c)$ is the 
interpolation function, which is also known as the weight function, from the particle's location $\mathbf{x}_p^{n+\frac{1}{2}}$
to the cell center $\mathbf{x}_c$. We note that a macro-particle represents millions of physical particles that 
are close to each other in the phase space, and each macro-particle may carry different amounts of charge corresponding to $q_p$ but
the charge per mass ratio is the same for all particles representing the same species (for example electrons).

At the end of one computational cycle, the particle's position and the electric field are at different stages. 
In order to evaluate and fix the error of Gauss's law at time stage $n+1$, we interpolate 
the charge density $\rho_c^{n+1}$ from $\rho_c^{n+\frac{3}{2}}$ and $\rho_c^{n+\frac{1}{2}}$.
The goal is to add a displacement $\Delta \mathbf{x}_p$ to each particle's position
$\tilde{\mathbf{x}}_p^{n+\frac{3}{2}}$ so that the density $\rho_c^{n+1}$ satisfies Gauss's law:
\begin{linenomath}
 \begin{equation}
\rho^{n+1}_c= \gamma\sum_p q_pW(\tilde{\mathbf{x}}_p^{n+\frac{3}{2}}+\Delta\mathbf{x}_p-\mathbf{x}_c) + (1-\gamma)\rho_c^{n+\frac{1}{2}} =
\frac{1}{4\pi} \nabla \cdot \mathbf{E}^{n+1},
\label{eq:gauss1}
\end{equation}
\end{linenomath}
where $\gamma$ is an interpolation coefficient. When $\gamma=0.5$, the interpolation 
is second-order accurate. But our tests suggest that using $\gamma=0.5$ may cause numerical oscillations. Similarly to the optimal choice of the $\theta$ parameter, we find that  
 $\gamma=0.51$ works very well. It sacrifices the accuracy slightly but eliminates the artificial oscillations. $\gamma = 0.51$ is used in this paper. Our goal is to displace the particles so that the equation above is satisfied
at all cell centers. This equation system is likely to be under-determined in general, 
because there are usually 
more particles (and corresponding unknown displacement vectors $\Delta \mathbf{x}_p$) than the number of cell centers (corresponding to the number of equations). The position correction can be applied 
to only one species (for example electrons only) or all species. 
In the following derivation of this accurate correction method, 
we assume that the correction is applied to all species. 

The displacement $\Delta \mathbf{x}_p$ should be small with respect to the cell size. 
Under the assumption of small displacements, the computation can be simplified
by linearizing the interpolation function:
\begin{linenomath}
 \begin{equation}
W(\tilde{\mathbf{x}}_p^{n+\frac{3}{2}}+\Delta\mathbf{x}_p-\mathbf{x}_c) = W(\tilde{\mathbf{x}}_p^{n+\frac{3}{2}}-\mathbf{x}_c) + \nabla W(\tilde{\mathbf{x}}_p^{n+\frac{3}{2}} - \mathbf{x}_c) \cdot \Delta \mathbf{x}_p + O((\Delta x)^2).
\label{eq:w-linear}
\end{equation}
\end{linenomath}
In our GL-ECSIM code, we use the zeroth order B-spline function $b_0$
(see Figure.\ref{fig:b-splines}a) to form the 3-dimensional shape function of the macro-particles:
\begin{linenomath}
 \begin{equation}
S(\mathbf{x}_p-\mathbf{x}_c) = \frac{1}{\Delta x \Delta y \Delta z} b_{0}
\Big(\frac{x_p-x_c}{\Delta x}\Big)b_{0}\Big(\frac{y_p-y_c}{\Delta y}\Big)b_{0}\Big(\frac{z_p-z_c}{\Delta z}\Big) .
\label{eq:s-define}
\end{equation}
\end{linenomath}
The $S$ function is a top-hat function centered around the particle with the width of the cell size. The interpolation function 
from a particle to a cell center is the integral of the particle's shape function over this cell, 
which leads to the first-order B-spline function $b_1$ (see Figure.\ref{fig:b-splines}b). In a 
three dimensions (3D), the interpolation function is
\begin{linenomath}
 \begin{equation}
W(\mathbf{x}_p-\mathbf{x}_c) = b_{1}\Big(\frac{x_p-x_c}{\Delta x}\Big)b_{1}\Big(\frac{y_p-y_c}{\Delta y}\Big)b_{1}\Big(\frac{z_p-z_c}{\Delta z}\Big)
\label{eq:w-define}
\end{equation}
\end{linenomath}
The $b_{1}\Big(\frac{x_p-x_c}{\Delta x}\Big)$ function is differentiable with respect to $x_p$ when 
$\frac{x_p-x_c}{\Delta x} \neq 0, \pm 1 $ (see Figure.\ref{fig:b-splines}c):
\begin{linenomath}
\begin{equation} \label{eq:db1}
\begin{array}{lc}
b_{1}^{'}(\frac{x_p-x_c}{\Delta x}) =\left\{
\begin{array}{lc c}
-1/\Delta x, & $if $ x_c<x_p<x_c+\Delta x\\
 1/\Delta x,  & $if $ x_c-\Delta x<x_p<x_c\\
0, & $if $ x_p<x_c-\Delta x $ or $ x_p>x_c+\Delta x .
 \end{array}
 \right. \\
\end{array}
\end{equation}
\end{linenomath}
This spatial derivative suggests that if we move a particle toward (away from) the cell center, the
interpolation weight from the particle to this cell center will increase (decrease). If the particle 
is so close to the cell center that the displacement $\Delta x_p$ makes the particle cross the cell 
center, we cannot predict the change of the interpolation weight from the $b_1$ derivative because 
the $b_1$ function is not differentiable at $b_1(0)$. For these particles, the linearization of
eq.(\ref{eq:w-linear}) is not valid. In practice, only a small portion of all the particles may 
encounter this problem when the displacement is generally small. This means that the 
non-differentiability will have little effect in general and the problem is getting smaller with smaller displacements. 

With the spatial derivative of the $b_1$ function known, the gradient of the interpolation function can be obtained. 
For example, when $x_c<x_p<x_c+\Delta x$, $y_c<y_p<y_c+\Delta y$ and $z_c<z_p<z_c+\Delta z$, the interpolation function is:
 \begin{linenomath}
 \begin{equation}
W(\mathbf{x}_p-\mathbf{x}_c) = \frac{(x_c+\Delta x-x_p)(y_c+\Delta y-y_p )(z_c+\Delta z-z_p )}{\Delta x \Delta y \Delta z} 
\label{eq:w-example}
\end{equation}
\end{linenomath}
and its gradient is:
 \begin{linenomath}
 \begin{equation}
\nabla W(\mathbf{x}_p-\mathbf{x}_c) = \left(
\frac{-W(\mathbf{x}_p-\mathbf{x}_c) }{x_c+\Delta x-x_p}, 
\frac{-W(\mathbf{x}_p-\mathbf{x}_c) }{ y_c+\Delta y-y_p}, 
\frac{-W(\mathbf{x}_p-\mathbf{x}_c) }{z_c+\Delta z-z_p}\right). 
\label{eq:dw}
\end{equation}
\end{linenomath}
From this example, we can see that the interpolation function is not linear and the 
$O((\Delta x)^2)$ term in eq.(\ref{eq:w-linear}) will not vanish.

We substitute eq.(\ref{eq:w-linear}) into eq.(\ref{eq:gauss1}) and drop the $O((\Delta x)^2)$ term to obtain the linearized Gauss's law constrain for a given cell center:
\begin{linenomath}
 \begin{equation}
g_c(\Delta \mathbf{x}_p) := \sum_p q_p \nabla W(\tilde{\mathbf{x}}_p^{n+\frac{3}{2}} - 
\mathbf{x}_c) \cdot \Delta \mathbf{x}_p - S_c = 0  \label{eq:dxp2}
\end{equation}
\end{linenomath}
where the constant term (independent of $\Delta\mathbf{x}_p$) is
\begin{linenomath}
\begin{equation}
S_c := \frac{1}{\gamma}\left[\frac{1}{4\pi} \nabla \cdot \mathbf{E}^{n+1} 
- \big( (1-\gamma) \rho_c^{n+\frac{1}{2}} 
+ \gamma \sum_p q_p W(\tilde{\mathbf{x}}_p^{n+\frac{3}{2}} - \mathbf{x}_c) \big) \label{eq:sc}\right]
\end{equation}
\end{linenomath}
Both $g_c(\Delta \mathbf{x}_p)$ and $S_c$ are defined at every cell center. 
To find a solution for the under-determined equations above while minimizing the displacements, we use the 
Lagrange multiplier method. The function we are trying to minimize is defined as
\begin{linenomath}
 \begin{equation}
f(\Delta \mathbf{x}_p) = \sum_p \frac{1}{2} (\Delta \mathbf{x}_p)^2 |q_p|^{\alpha}
\label{eq:f-func}
\end{equation}
\end{linenomath}
where $\alpha$ is a non-negative exponent to be specified later. 
Our goal is to minimize the function $f(\Delta \mathbf{x}_p)$ provided that eq.(\ref{eq:dxp2}) is
satisfied for each cell center. The Lagrange function is:
\begin{linenomath}
 \begin{equation}
 \begin{split}
L(\Delta \mathbf{x}_p, \lambda_c) & = f(\Delta \mathbf{x}_p) - \sum_c \lambda_c g_c(\Delta \mathbf{x}_p) \\
 & = \sum_p \frac{1}{2} (\Delta \mathbf{x}_p)^2 |q_p|^{\alpha} - \sum_c \lambda_c 
 \Big[ \sum_p   q_p \nabla W(\tilde{\mathbf{x}}_p^{n+\frac{3}{2}} - \mathbf{x}_c) \cdot \Delta \mathbf{x}_p - S_c \Big]
\end{split}
\label{eq:L-func}
\end{equation}
\end{linenomath}
where $\lambda_c$ are the Lagrange multiplier for all the cell centers. 
The function $f$ reaches a local extrema if 
the Lagrange function's partial derivatives with respect to the displacements $\Delta \mathbf{x}_p$ and 
the Lagrange multipliers $\lambda _c$ are all zero:
\begin{linenomath}
 \begin{eqnarray}
\frac{\partial L}{\partial \lambda_c}  = g_c(\Delta \mathbf{x}_p) = 
\sum_p q_p \nabla W(\tilde{\mathbf{x}}_p^{n+\frac{3}{2}} - \mathbf{x}_c) \cdot \Delta \mathbf{x}_p - S_c = 0 
\label{eq:dl_dlambda} \\ 
\frac{\partial L}{\partial \Delta \mathbf{x}_p}  = \Delta \mathbf{x}_p |q_p|^{\alpha}  - 
 \sum_c \lambda_c q_p \nabla W(\tilde{\mathbf{x}}_p^{n+\frac{3}{2}} - \mathbf{x}_c) = 0 .  \label{eq:dl_ddxp}
 \end{eqnarray}
\end{linenomath}
Thanks to the linearization, the displacement of each particle can be easily expressed as a function of $\lambda_c$ by solving eq.(\ref{eq:dl_ddxp}):
\begin{linenomath}
 \begin{equation}
 \Delta \mathbf{x}_p  = 
 \sum_c \lambda_c |q_p|^{-\alpha} q_p \nabla W(\tilde{\mathbf{x}}_p^{n+\frac{3}{2}} - \mathbf{x}_c) \label{eq:dxp}
 \end{equation}
\end{linenomath}
and substituted into eq.(\ref{eq:dl_dlambda}) to obtain a linear system of equations that only contains $\lambda_c$ as unknowns:
 \begin{linenomath}
 \begin{eqnarray}
 \frac{\partial L}{\partial \lambda_c} =  \sum_p q_p \nabla W(\tilde{\mathbf{x}}_p^{n+\frac{3}{2}} 
 - \mathbf{x}_c) \cdot  \Big[|q_p|^{-\alpha} q_p\sum_{c'} \lambda_{c'}  \nabla W(\tilde{\mathbf{x}}_p^{n+\frac{3}{2}} 
 - \mathbf{x}_{c'}) \Big]  - S_c = 0 .\label{eq:lambda1} 
 \end{eqnarray}
\end{linenomath}
We note that this is an equation for cell center $c$ so we introduced $c'$ for the summation. After exchanging the order of the two summations for $c'$ and $p$, we obtain
 \begin{linenomath}
 \begin{eqnarray}
\sum_{c'} M_{cc'}  \lambda_{c'} = S_c \label{eq:lambda2} 
 \end{eqnarray}
\end{linenomath}
where the matrix element $M_{cc'}$ is defined as:
 \begin{linenomath}
 \begin{eqnarray}
M_{cc'} := \sum_p |q_p|^{2-\alpha} \nabla W(\tilde{\mathbf{x}}_p^{n+\frac{3}{2}} - 
\mathbf{x}_c) \cdot \nabla W(\tilde{\mathbf{x}}_p^{n+\frac{3}{2}} - \mathbf{x}_{c'}).\label{eq:M} 
 \end{eqnarray}
\end{linenomath}
Once the `mass matrix' $M$ is calculated, the Lagrange multipliers $\lambda_c$ can be obtained by solving the linear 
system eq.(\ref{eq:lambda2}), then we can calculate the particle displacement $\Delta \mathbf{x}_p$ from
eq.(\ref{eq:dl_ddxp}) and add the displacements to $\tilde{\mathbf{x}}_p^{n+\frac{3}{2}}$ to obtain the corrected
particle positions:
 \begin{linenomath}
 \begin{eqnarray}
\mathbf{x}_p^{n+\frac{3}{2}} =  \tilde{\mathbf{x}}_p^{n+\frac{3}{2}} + \Delta \mathbf{x}_p .
\label{eq:x_corrected}
 \end{eqnarray}
\end{linenomath}
We use the GMRES iterative method to solve eq. (\ref{eq:lambda2}). 

Since the $O((\Delta x)^2)$ term is not zero in eq.(\ref{eq:w-linear}), there is still an 
error of $O((\Delta x)^2)$ in Gauss's law (see eq.(\ref{eq:gauss1}) after the correction. 
To further minimize the error, we can repeat the correction several times. The particle 
displacement decreases when we repeat the correction, so it also helps to reduce the influence 
of the singularity in the $b_1$ derivative (see eq.(\ref{eq:db1})). In section \ref{section:test},
we show that after three corrections, the error in Gauss's law reduces to a very small value. 

We can now determine the most sensible value for the $\alpha$ exponent introduced in 
eq.(\ref{eq:f-func}). If two particles of
the same species overlap with each other before the correction, it is natural to 
correct them with the same displacement, i.e., their displacements $\Delta \mathbf{x}_p$ should
not depend on the particle's charge $q_p$. According to eq.(\ref{eq:dxp}) this will hold 
if we set $\alpha=1$, which is the value used in all simulations in this paper. 
When $\alpha = 1$, eq.(\ref{eq:f-func}) implies that
the Lagrange function minimizes the sum of $|q_p|(\Delta x_p)^2$ over the particles.

We assumed that 
all species are corrected above, but we have the freedom to correct one species only.
In that case only the particles that require correction are looped through to calculate the matrix $M$ (see eq.(\ref{eq:M})) 
and the displacement $\Delta \mathbf{x}_p$ (see eq.(\ref{eq:dxp})), which are the two most expensive 
parts of one correction cycle. So it is better to correct only one species in terms of
computational efficiency. We find that correcting the lightest species (typically electrons) 
only is a reasonable choice in practice.

\subsubsection{The approximate global correction}
The accurate correction reduces the error in Gauss's law to the iterative tolerance level. But 
it requires looping through particles to calculate the matrix $M$ (see eq.(\ref{eq:M})). This 
step is computationally expensive. If the goal is to suppress the growth of the error in
Gauss's law instead of eliminating it entirely, the calculation of the matrix $M$ can be avoided. 

Boris' electric field correction method solves the following Poisson equation of the scalar 
function $\phi$ defined at cell centers:
 \begin{linenomath}
 \begin{eqnarray}
\nabla^2 \phi = \nabla \cdot \mathbf{{\tilde E}}^{n+1} - 4\pi \tilde \rho_c^{n+1} ,
\label{eq:phi1}
 \end{eqnarray}
\end{linenomath}
where $\mathbf{\tilde E}$ and $\tilde \rho_c$ are the uncorrected electric field and charge density at the cell center. After  $\phi$ is obtained, the electric field is corrected to satisfy Gauss's law:
 \begin{linenomath}
 \begin{eqnarray}
\mathbf{E}^{n+1} = \mathbf{\tilde E}^{n+1} - \nabla \phi .
\label{eq:E-boris}
 \end{eqnarray}
\end{linenomath}
 
Instead of correcting the electric field, we design an analogous algorithm that corrects the particle positions. Similar to the Boris field correction, we solve the Poisson equation (\ref{eq:phi1}) first with the GMRES scheme. The charge density is interpolated as
 \begin{linenomath}
 \begin{eqnarray}
\tilde \rho_c^{n+1} = \gamma \tilde{\rho}_c^{n+\frac{3}{2}} + (1-\gamma)\rho_c^{n+\frac{1}{2}}
\label{eq:rhoc-approximate}
 \end{eqnarray}
\end{linenomath}
where the tilde represents the charge density before position correction and $\gamma=0.51$ is an interpolation coefficient as 
in eq.(\ref{eq:gauss1}). If we could find displacements $\Delta \mathbf{x}_p$ for each 
particle so that
 \begin{linenomath}
 \begin{eqnarray}
\rho^{n+\frac{3}{2}}(\mathbf{\tilde{x}}_p^{n+\frac{3}{2}}+\Delta \mathbf{x}_p) = 
\tilde{\rho}^{n+\frac{3}{2}}(\mathbf{\tilde{x}}_p^{n+\frac{3}{2}}) + \frac{1}{4\pi \gamma} \nabla ^2 \phi,
\label{eq:rho-goal}
 \end{eqnarray}
\end{linenomath}
it is easy to show that the corrected and time interpolated charge density  $\rho_c^{n+1}$ and 
the original electric field $\mathbf{E}^{n+1}=\mathbf{\tilde E}^{n+1}$ will satisfy Gauss's law. So, the goal is to find the displacements 
$\Delta \mathbf{x}_p$ that satisfy the above equation where $\phi$ is given by eq.(\ref{eq:phi1}).

When we add the displacement $\Delta \mathbf{x}_p$ to a particle, it is equivalent to add a `virtual current' $\mathbf{j}_v$ for a `virtual time' $\Delta t_v$ to change the charge density 
from $\tilde{\rho}^{n+\frac{3}{2}}(\mathbf{\tilde{x}}_p^{n+\frac{3}{2}}) $ to 
$ \rho^{n+\frac{3}{2}}(\mathbf{\tilde{x}}_p^{n+\frac{3}{2}}+\Delta \mathbf{x}_p)$.
The charge conservation equation describes how the `virtual current' changes the charge density:
 \begin{linenomath}
 \begin{eqnarray}
\rho^{n+\frac{3}{2}}(\mathbf{\tilde{x}}_p^{n+\frac{3}{2}}+\Delta \mathbf{x}_p) - 
\tilde{\rho}^{n+\frac{3}{2}}(\mathbf{\tilde{x}}_p^{n+\frac{3}{2}})
= \nabla \cdot (\Delta t_v  \mathbf{j}_v) + \text{discretization error}.
\label{eq:drho}
 \end{eqnarray}
\end{linenomath}
Combining eq.(\ref{eq:drho}) and eq.(\ref{eq:rho-goal}), we obtain the equation for 
the $ \Delta t_v  \mathbf{j}_v$ term:
 \begin{linenomath}
 \begin{eqnarray}
 \Delta t_v  \mathbf{j}_v = \frac{1}{4\pi \gamma} \nabla \phi + \text{discretization error}. 
\label{eq:jv}
 \end{eqnarray}
\end{linenomath}
For the sake of simplicity, we only displace the electrons or the lightest species to 
create the `virtual current'. For a given position $\mathbf{x}_p$, if we displace the surrounding
electrons by $\Delta \mathbf{x}_p$, it will generate a 'virtual current':
 \begin{linenomath}
 \begin{eqnarray}
 (\Delta t_v  \mathbf{j}_v)_{p} = \rho_{e,p} \Delta \mathbf{x}_p \approx \rho_{e,g} \Delta \mathbf{x}_p
\label{eq:jv-p}
 \end{eqnarray}
\end{linenomath}
where $\rho_{e,p}$, $\rho_{e,g}$ are the electron charge densities at $\mathbf{x}_p$ and its closest node, respectively. 
Combining the two equations above and ignoring the discretization errors, the displacement $\Delta\mathbf{x}_p$ is obtained as
 \begin{linenomath}
 \begin{eqnarray}
 \Delta \mathbf{x}_p = \frac{1}{4\pi \gamma\rho_{e,g}} \nabla \phi.
\label{eq:jv-xp}
 \end{eqnarray}
\end{linenomath} 

This global approximate correction method solves a Poisson's equation to distribute the 'virtual current'
globally. It does not eliminate the error in Gauss's law exactly, but it pushes the particles toward 
the direction to reduce the error. To avoid potential overshoot, we can apply partial correction only:
 \begin{linenomath}
 \begin{eqnarray}
 \Delta \mathbf{x}_p =   \frac{\epsilon}{4\pi \gamma\rho_{e,g}} \nabla \phi  . 
\label{eq:jv-xp-eps}
 \end{eqnarray}
\end{linenomath}
where $\epsilon$ is a constant between 0 and 1. We use $\epsilon=0.9$ in practice. The spatial
discretization is described in the section \ref{section:approximate-local}.

\subsubsection{The approximate local correction}
\label{section:approximate-local}
\begin{figure}
\begin{center}
\includegraphics[width=0.5\textwidth]{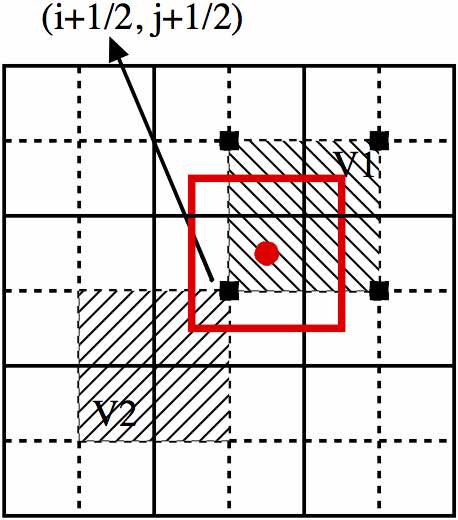}
\caption{The black solid lines represent the cell edges. The black squares are the cell centers. 
The red square represents the shape function $S_p$ of a macro-particle with its position $\mathbf{x}_p$ marked by the red circle. The two 
shaded squares are two complementary volumes (node-centered volumes) V1 and V2. 
\label{fig:mesh}
}
\end{center}
\end{figure}

The approximate global correction method described in the previous section needs to solve a Poisson equation. 
Its computational cost is negligible within our GL-ECSIM scheme. But the cost may not be acceptable 
for an explicit PIC algorithm. To avoid solving the Poisson equation, we introduce a local correction 
method.

Again, we only correct the electrons for simplicity. We calculate the relative error at each 
cell center first:
 \begin{linenomath}
 \begin{eqnarray}
r_c = \frac{\tilde\rho_c^{n+1} -\nabla \cdot \mathbf{{E}}^{n+1}/(4\pi)}{\gamma \rho_{e,c}}
 \end{eqnarray}
\end{linenomath}
where $\tilde \rho_c^{n+1}$ is obtained from eq.(\ref{eq:rhoc-approximate}). The displacement 
$\Delta \mathbf{x}_p$ for a particle at $\mathbf{x}_p$ is calculated from
 \begin{linenomath}
 \begin{eqnarray}
(\Delta x_p/\Delta x, \Delta y_p/\Delta y, \Delta z_p/\Delta z) = 
-\epsilon \left(
\frac{\Delta x}{2} \frac{\partial r_c}{\partial x},
\frac{\Delta y}{2} \frac{\partial r_c}{\partial y}, 
\frac{\Delta z}{2}  \frac{\partial r_c}{\partial z}
\right)_p
\label{eq:dx-local-correction}
 \end{eqnarray}
\end{linenomath}
where the right-hand side is the difference of the relative error $r_c$ in the three directions, 
$\Delta x$, $\Delta y$ and $\Delta z$ are the cell sizes,  and $\epsilon$ is the correction 
ratio between 0 and 1. The difference of the relative error $r_c$ indicates the direction to move 
particles. As an example, let us consider a uniform 1D simulation with a pair of electron 
and ion at each node at time stage $n+\frac{1}{2}$. Assume the cell size is 1, each ion macro-particle has 
charge $q_i$ and each electron has charge $q_e=-q_i$, so the cell center electron charge 
density is $\rho_{e,c} = q_e$ and the net charge at $n+\frac{1}{2}$ stage is zero. We assume the 
electric field at $n+1$ is also zero. If an electron macro-particle at the cell center $x_{i}$ is 
misplaced at $x_{i} + 0.1\Delta x$ at $n+\frac{3}{2}$ stage and other particles do not move, the electron charge
at cell centers $x_{i-1/2}$ and $x_{i+1/2}$ will become $0.9q_e$ and $1.1q_e$, respectively. The relative
errors $r_c$ at $x_{i-1/2}$ and $x_{i+1/2}$ are $\frac{-0.1}{0.9}\approx -0.11$ and $\frac{0.1}{1.1}\approx 0.091$, 
respectively. Based on the correction formula above, the correction for this electron particle is 
$ \Delta x_p/\Delta x = - \frac{\Delta x}{2} \frac{\partial r_c}{\partial x} \approx 
 -(0.091+0.11)/2 \approx -0.1$ when $\epsilon=1$, which means the electron at $x_i + 0.1\Delta x$ will 
 be moved back to $x_i$. For this simple example, $\epsilon=1$ cancels the error almost perfectly. 
 
Figure \ref{fig:mesh} shows a two-dimensional example. Among the 4 cell centers around the particle in the 
figure, the smallest index cell center is $(i+\frac{1}{2},j+\frac{1}{2})$. Based on the relative errors at 
these 4 cell centers, this particle will move toward or away from the cell center $(i+\frac{1}{2},j+\frac{1}{2})$. 
However, the information in the complementary 
volume V2 has no influence on this particle although particles inside V2 also contribute to cell center 
$(i+\frac{1}{2},j+\frac{1}{2})$. Due to the locality of this correction method, it is impossible to find 
a correction ratio $\epsilon$ to eliminate the error accurately in general. A large $\epsilon$
can lead to overshoots easily, while a small $\epsilon$ may not be sufficient to suppress the growth of the error. 
Our tests suggest that $\epsilon=0.5$ reaches a reasonable balance between the effectiveness and robustness, and it is
used in the following numerical tests. 

We use Figure \ref{fig:mesh} to illustrate the calculation of the spatial derivatives in 
eq. (\ref{eq:dx-local-correction}) and eq. (\ref{eq:jv-xp-eps}). Assume the particle 
is at $(x_p, y_p)$ and we need to calculate $\frac{\partial r_c}{\partial x}$. We interpolate
$r_{x_{i+1/2}, y_p}$ ($r_{x_{i+3/2}, y_p}$) from $r_{x_{i+1/2}, y_{i+1/2}}$ and 
$r_{x_{i+1/2}, y_{i+3/2}}$ ($r_{x_{i+3/2}, y_{i+1/2}}$ and $r_{x_{i+3/2}, y_{i+3/2}}$) first. 
Then the spatial derivative is obtained by 
$\frac{\partial r_c}{\partial x} = (r_{x_{i+3/2}, y_p} - r_{x_{i+1/2}, y_p})/\Delta x$. 

\subsubsection{Limiting the displacement}
All the three correction methods described above assume that if a particle moves toward (away from)
a cell center, its charge contribution to this center would increase (decrease). This assumption 
is true only when the particle center does not cross the complementary volume boundaries. When
the displacement is small, there are not too many particles violating this assumption and the correction
methods work well. However, in the region where the plasma is rarefied or the numerical error in Gauss's law 
is large, the displacement can be large compared to the cell size. To fix this problem, 
we limit the displacement with the following simple algorithm:
 \begin{linenomath}
 \begin{eqnarray}
\Delta \mathbf{x}_p^{new} = \text{min}\left(1,c_0\frac{\Delta x}{|\Delta \mathbf{x}_p |}\right)\,\Delta \mathbf{x}_p
\label{eq:dx-limit}
 \end{eqnarray}
\end{linenomath}
where $\Delta \mathbf{x}_p$ the particle displacement calculated by one of the correction methods, 
$\Delta x$ is the cell size in the x-direction, and $c_0$ is the maximum allowed relative displacement.
We use $c_0=0.1$ for the simulations. 

\subsection{Spatial discretization}
\label{section:spatial-dis}
The spatial discretization of the semi-discretized equations eq.(\ref{eq:E-pse}) and eq.(\ref{eq:theta-B}) 
on a uniform Cartesian grid can be done following the iPIC3D convention. Since $\mathbf{E}$ and $\mathbf{B}$ 
are staggered in space, we need first-order derivatives from cell centers to nodes and from nodes to 
cell centers, and second-order derivatives from nodes to nodes. The node-to-node second-order derivatives can 
be obtained in two steps: first calculate the node-to-center first-order derivatives and then calculate the 
center-to-node derivatives of these first-order derivatives. Each cell center (node) first-order derivative 
is calculated by averaging the 4 nodes (centers) in the transverse directions and then taking the difference 
between the two averaged values along the direction of the derivative. For example, the cell centered 
first-order derivative of $E_x$ in the $x$ direction is calculated as
\begin{linenomath}\begin{equation}
\begin{split}
\left. \frac{\partial E_x}{\partial x}\right|_{i+\frac{1}{2},j+\frac{1}{2},k+\frac{1}{2}} = 
\frac{1}{\Delta x}\sum _{l,m=0}^{l,m=1}\frac{1}{4} (E_{x,i+1,j+l,k+m} - E_{x,i,j+l,k+m})
\end{split}
\label{eq:first-derivative-compact}
\end{equation}\end{linenomath}
where the integer indices i, j and k represent the cell nodes while the half indices represent the cell centers. 
All the spatial derivatives in eq.(\ref{eq:E-pse}) can be calculated based on this rule. 
We note that not all spatial discretizations satisfy the identities needed
for energy conservation \cite{Lapenta:2017a} but, fortunately, the 
discretization described above does. It also satisfies the identity $\nabla\times\nabla\times = (\nabla\nabla\cdot) - \nabla^2$ used in deriving 
eq.(\ref{eq:ECSIM-E}).

This finite difference algorithm uses as few neighbors as possible while maintaining 
symmetric discrete formulas that satisfy the various identities.
It is quite optimal and it behaves well for most of our simulations. But spurious 
short-wavelength oscillations with wavelength of $\sim 2$ cells may occur with 
this compact discretization for some simulations. We found 
that using an extended stencil for part of the the spatial discretization of 
$\nabla \cdot \mathbf{E}^{n+\theta}$ in 
eq.(\ref{eq:ECSIM-E}) helps to suppress these oscillations. We take
$\partial E_x/\partial x$ at the cell center as an example to define the
difference formula with an extended stencil:
\begin{linenomath}\begin{equation}
\begin{split}
\left. \frac{\partial E_x}{\partial x}\right|_{i+\frac{1}{2},j+\frac{1}{2},k+\frac{1}{2}} = 
\frac{1}{2\Delta x}\sum _{l,m=-1}^{l,m=1}\frac{1}{9}(E_{x,i+\frac{3}{2},j+\frac{1}{2}+l,k+\frac{1}{2}+m} -
 E_{x,i-\frac{3}{2},j+\frac{1}{2}+l,k+\frac{1}{2}+m})
\end{split}
\label{eq:first-derivative-large}
\end{equation}\end{linenomath}
where the cell center electric field values, such as $E_{x,i+\frac{3}{2},j+\frac{1}{2},k+\frac{1}{2}}$,
 are averaged form the nearby 8 nodes. We denote the
divergence calculated on the extended stencil shown by eq.(\ref{eq:first-derivative-large}) 
as $\nabla' \cdot \mathbf{E}^{n+\theta}$, while $\nabla \cdot \mathbf{E}^{n+\theta}$ represents the usual compact discretization of
 eq.(\ref{eq:first-derivative-compact}).
The difference of these two divergence
operators can be used to diffuse the
oscillatory errors related to the $\nabla\nabla\cdot\mathbf{E}$ term.
Using a linear combination of $\nabla' \cdot \mathbf{E}^{n+\theta}$ and $\nabla \cdot \mathbf{E}^{n+\theta}$, the electric field equation becomes: 
\begin{linenomath}\begin{eqnarray}
\mathbf{E}^{n+\theta} + \delta^2 \left[ \nabla(c_{cpt}\nabla \cdot \mathbf{E}^{n+\theta} + (1-c_{cpt}) \nabla'\cdot\mathbf{E}^{n+\theta})) - \nabla ^2 \mathbf{E}^{n+\theta}\right] = \\
\mathbf{E}^{n} + \delta\left(\nabla \times \mathbf{B}^n 
- \frac{4\pi}{c}\mathbf{\bar{J}}\right), \nonumber
\label{eq:E-cpt}
\end{eqnarray}\end{linenomath}
where the coefficient $c_{cpt}$ is the fraction of the divergence calculated with the compact derivative.

We illustrate the smoothing effect of using the extended stencil for the divergence operator with  
a 1D example. Let us assume that there is charge separation in a 1D simulation along the x-direction 
that generates a variation in the $E_x$ component with a short wavelength. Since $\nabla \times \mathbf{E}=0$ for this case, 
the $\nabla( \nabla \cdot \mathbf{E}^{n+\theta})$
and $\nabla ^2 \mathbf{E}^{n+\theta}$ terms cancel each other both analytically and numerically when
the compact derivatives are applied. However, if $c_{cpt}$ is less than 1 so that the extended stencil
derivative $\nabla'\cdot$ is also used, then in effect we add
\begin{equation}
\delta^2(1-c_{cpt})\nabla(\nabla-\nabla') E^{n+\theta}_x
\label{eq:E-cpt-diffusion}
\end{equation}
to the right hand side of the original equation (\ref{eq:ECSIM-E}). The leading term in the Taylor expansion will
be a fourth derivative $-\delta^2(1-c_{cpt}) \frac{\Delta x ^2}{2} \frac{\partial^4E_x}{\partial x^4}$ 
since the third derivative has zero coefficient due to the symmetry of the discrete 
divergence and gradient operators. 
This operator has a net effect of smoothing the short-wavelength oscillations 
in $E_x$. 

We remark that when $c_{cpt}$ is not 1, i.e., the extended stencil divergence of the electric field is used, the
total energy is not exactly conserved any more. In section \ref{section:test}, we are going to show that 
simulations with $c_{cpt}=0.9$ suppress the oscillations while still conserve energy reasonably well. 

\section{Numerical tests}
\label{section:test}

This section presents three numerical tests to demonstrate the performance of the GL-ECSIM algorithm.
The two-dimensional (2D) magnetosphere simulation and the 2D reconnection test show the improvement
of the GL-ECSIM scheme compared to iPIC3D and the original ECSIM algorithm. The 1D Weibel instability test 
demonstrates that the particle position correction step does not change the physics. 

Table~\ref{tb:parameters} shows 9 different parameter combinations for the tests. 
We performed simulations with iPIC3D, the original ECSIM and GL-ECSIM. For the original 
ECSIM, the role of $\theta$ is studied (ECSIM-1 and ECSIM-2), 
and we show that Marder's pseudo-current method \cite{Marder:1987} does no work well (ECSIM-3).
For GL-ECSIM, we show that the extended stencil spatial discretization helps to suppress 
the short-wavelength oscillations by comparing GL-ECSIM-1 and GL-ECSIM-2, and we also compare different particle position
correction methods (GL-ECSIM-2 to GL-ECSIM-5).

In this test section, we set the electric field solver tolerance to be $10^{-6}$.
For the accurate correction method, the correction procedure is repeated three times 
per computational cycle. A iteration tolerance of 0.01 and a maximum iteration number 
of 20 are used for the linear equation systems of the correction methods. We have not 
implemented any preconditioner for the iterative solver, so the linear equations may not 
be able to converge to the tolerance 0.01 within 20 iterations. But the correction
methods still work well as the following tests demonstrate.

\begin{table}[ht]
\caption{Simulation parameters and the normalized wall time for the 2D reconnection simulations.
In the particle correction method column, `accurate', `approximate-global' and `approximate-local' 
represent three methods described in section \ref{section:particle correction}, and `all' 
indicates that the correction method is applied to all species, otherwise the correction is 
applied to electrons only. $c_{cpt}$ is the coefficient of the compact $\nabla \cdot \mathbf{E}$ 
discretization. $c_{pc}$ is the coefficient of the pseudo-current term. The 2D magnetic 
reconnection (MR) simulation wall time is normalized by the iPIC3D wall time.  }
\centering
\begin{tabular}{c c c c c c }
\hline
 Simulation ID    & $\theta$ & Correction method & $c_{cpt}$  & $c_{pc}$ & MR wall time \\
\hline
iPIC3D         & 1.0      & N/A                 & N/A         & 1.0   & 1.0  \\
ECSIM-1        & 0.5      & N/A                 & 1           & 0     & 1.8  \\
ECSIM-2        & 0.51     & N/A                 & 1           & 0     & 1.8  \\
ECSIM-3        & 0.51     & N/A                 & 1           & 0.1   & N/A  \\
GL-ECSIM-1     & 0.51     & accurate            & 1           & 0     & 2.6  \\
GL-ECSIM-2     & 0.51     & accurate            & 0.9         & 0     & 2.6  \\
GL-ECSIM-3     & 0.51     & accurate-all        & 0.9         & 0     & 2.9  \\
GL-ECSIM-4     & 0.51     & approximate-global  & 0.9         & 0     & 2.1  \\
GL-ECSIM-5     & 0.51     & approximate-local   & 0.9         & 0     & 2.0  \\

\hline
\label{tb:parameters}
\end{tabular}
\end{table}


\subsection{Two-dimensional magnetosphere simulation}
The numerical modeling of the 3D magnetosphere has been the original motivation for us to develop the 
GL-ECSIM method. Here we use a 2D magnetosphere simulation to show the problems we encountered with 
iPIC3D and ECSIM, and also to demonstrate that GL-ECSIM cures these issues. 

In the 2D Earth's magnetosphere simulation, we solve the ideal MHD equations with a separate electron
 equation to capture the global structure of the 2D magnetosphere. After a steady solution is obtained,
 we use the embedded PIC model to cover Earth's dayside magnetopause. The MHD code and the PIC code are two-way
 coupled. More details about the MHD-EPIC algorithm can be found in \cite{Daldorff:2014,Chen:2017}
 
\begin{figure}
\begin{center}
\includegraphics[width=\textwidth]{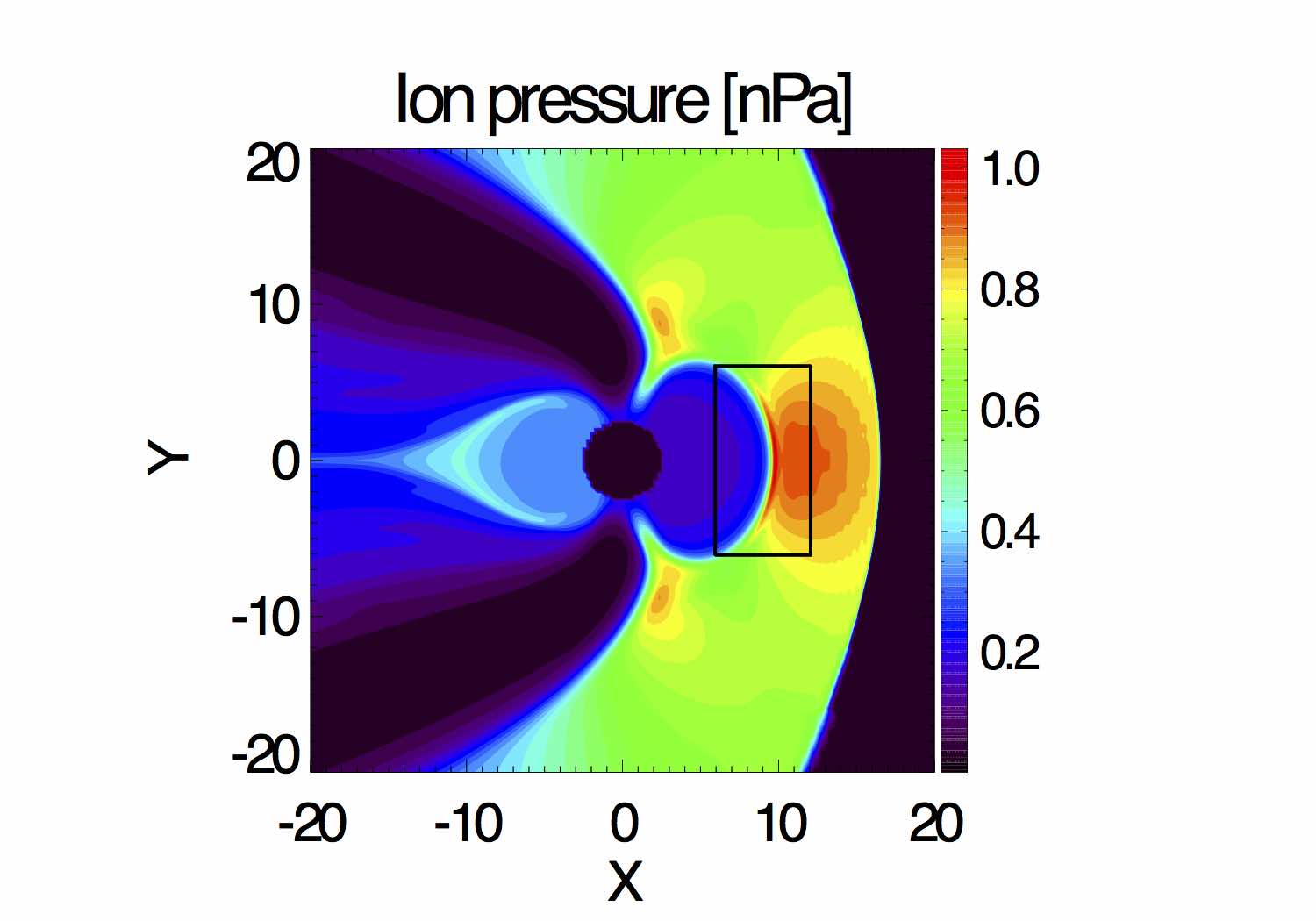}

\caption{The ion pressure of the 2D magnetosphere simulation. The region inside the black 
rectangle is simulated by the PIC code. 
\label{fig:pic_box}
}
\end{center}
\end{figure}

The 2D simulation domain extends from $x=-480\,R_E$ to $x=32\,R_E$ and $y=-128\,R_E$ to $y=128\,R_E$,
 where $R_E=6380$\,km is Earth's radius. The intrinsic magnetic field is represented by a 2D line dipole 
 with magnetic field strength -3110 nT at the magnetic equator. The dipole is aligned with the $Y$ axis. 
 The field strength of the 2D dipole is chosen so that the magnetopause forms at about the same 
 distance ($\approx 10\,R_E$) as in reality. The inner boundary condition is set at $r = 2.5\,R_E$ with 
 a fixed plasma density $10\,\text{amu/cc}$ and zero plasma velocity. The external magnetic 
 field (total field minus the intrinsic dipole) and the ion and electron pressures have zero gradient 
 inner boundary conditions. The solar wind enters the simulation domain from the $+x$ direction with 
 mass density $\rho_{mass}=5\,\text{amu/cc}$, electron pressure $p_e = 0.0124\,\text{nP}$, 
 ion pressure $p_i = 0.0062\,\text{nP}$, plasma velocity $\mathbf{u} = [-400,0,0]\,\text{km/s}$, 
 and magnetic field $\mathbf{B} = [-0.1, -0.5, 0]\,\text{nT}$. Figure~\ref{fig:pic_box} shows the ion pressure 
 in part of the simulation domain. After the MHD code reaches a steady state, the embedded PIC model is
 used to simulate the dayside reconnection region. The PIC region covers $6\,R_E < x < 12\,R_E$ and $-6\,R_E<y<6\,R_E$ 
 shown by the black box in Figure~\ref{fig:pic_box}. The ion mass-charge ratio $m_i/q_i$ is set to be 
 32 times larger than the ratio of a proton so that the ion inertial length $d_i$ is about $0.27\,R_E$ in the 
 magnetosheath (see \cite{Toth:2017} for more detail on the scaling). A reduced ion-electron mass 
 ratio $m_i/m_e=100$ is used so that the electron skin depth $d_e$ is about $d_i/10\approx 0.027\,R_E$.
The PIC code resolution is $1/32\,R_E$, so that there are about 10 cells per ion inertial length or 1 cell 
per electron skin depth. 400 macro-particles per cell per species are used. The time step is fixed to be 
$\Delta t=0.05\,s$ unless otherwise specified, and the corresponding CFL number 
$\text{CFL = max}(v_{x,e,th}^{max}/\Delta x,v_{y,e,th}^{max}/\Delta y, v_{z,e,th}^{max}/\Delta z )\Delta t $ is 
about 0.25, where $v_{e,th}^{max}$ is the maximum electron thermal velocity component. A reduced speed of 
light $c=3000\,\text{km/s}$ is used. 

\begin{figure}
\begin{center}
\includegraphics[width=1.0\textwidth]{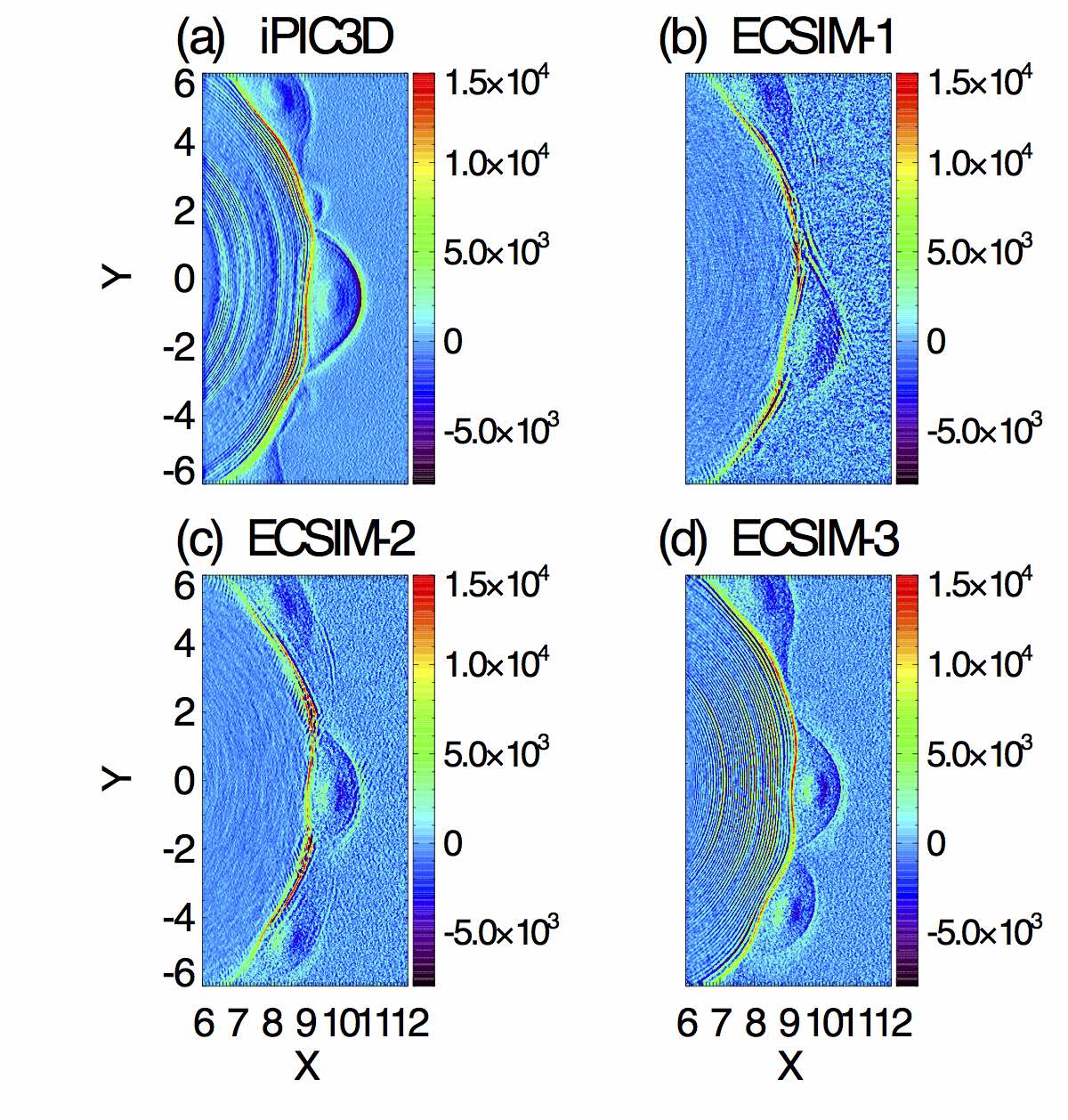}
\caption{The electric field $E_x[\text{nT\,km/s}]$ of the 2D magnetosphere simulations inside the PIC domain at $t=400s$ with four different simulation parameters described in Table \ref{tb:parameters}.
iPIC3D produces short-wavelength oscillations inside the magnetosphere. ECSIM with $\theta = 0.5$ (ECSIM-1)
generates more noise in the magnetosheath than ECSIM with $\theta = 0.5$ (ECSIM-2). There are some spurious 
small scale oscillations near the magnetopause and around the edge of a plasmoid for both ECSIM-1 and ECSIM-2. 
If the pseudo-current is used to fix the error in Gauss's law (ECSIM-3), it generates oscillations that 
are similar to the iPIC3D code.
\label{fig:mp_ex}
}
\end{center}
\end{figure}

\begin{figure}
\begin{center}
\includegraphics[width=0.8\textwidth]{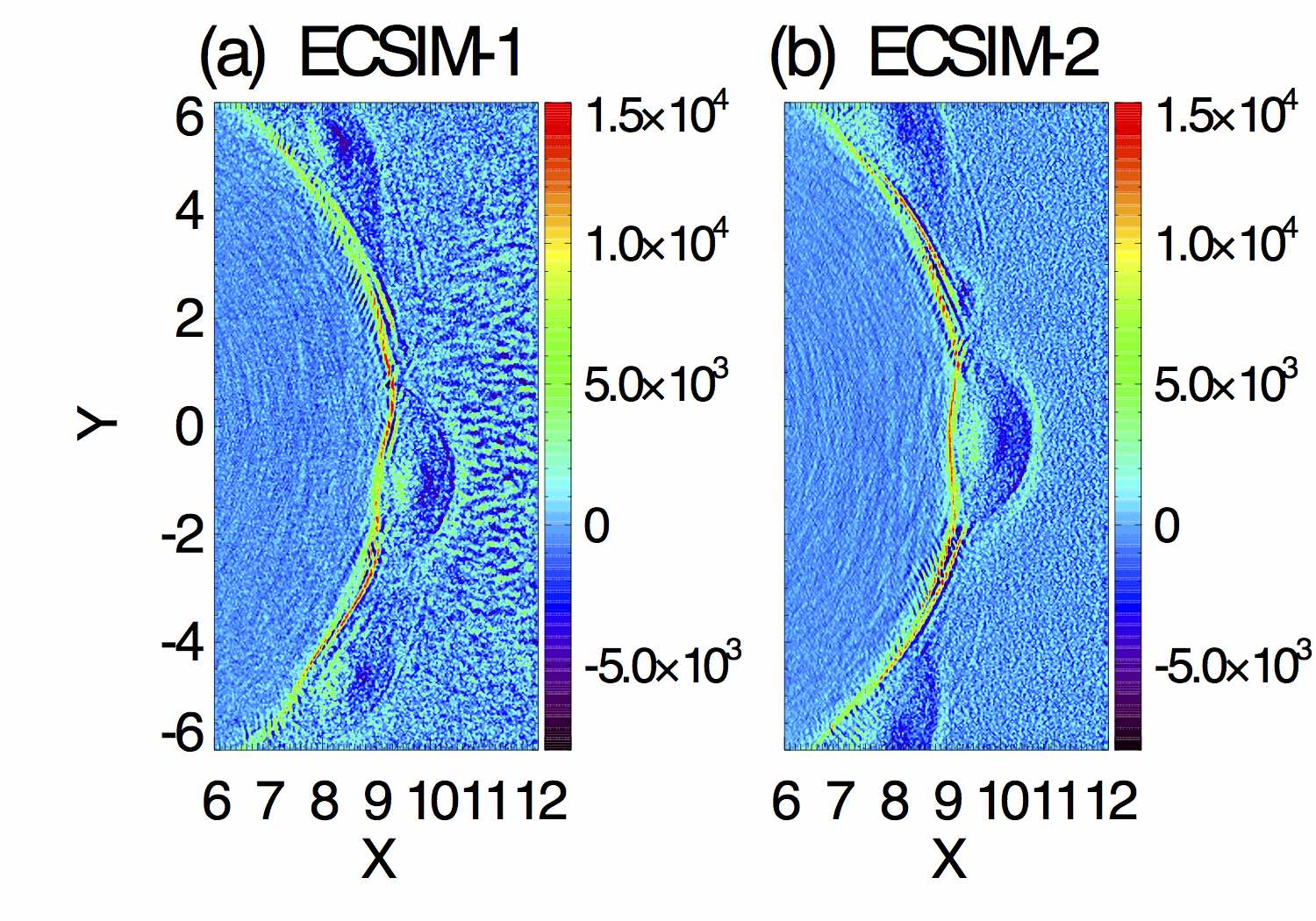}
\caption{$E_x[\text{nT\,km/s}]$ of simulations with $\Delta t = 0.1s$. ECSIM with 
$\theta = 0.5$ (ECSIM-1) produces wave-like structure in the magnetosheath, while ECSIM with 
$\theta = 0.51$ (ECSIM-2) does not. 
\label{fig:mp_ex_th}
}
\end{center}
\end{figure}

Figure~\ref{fig:mp_ex} compares the electric field component $E_x$ inside the PIC domain 
at $t=400\,s$ for iPIC3D, ECSIM with $\theta=0.5$ (ECSIM-1), ECSIM with $\theta = 0.51$ (ECSIM-2), 
and ECSIM with $\theta=0.51$ and the pseudo-current term (ECSIM-3). iPIC3D produces short-wavelength
oscillations in the magnetosphere. Our numerical tests show its wavelength is proportional to the cell 
size, so the oscillations can not be physical. The oscillations can be reduced by smoothing the electric
field after each update \cite{Toth:2017}. ECSIM-1 and ECSIM-2 successfully suppress the magnetosphere 
oscillations, but there are some spurious small scale oscillations near the magnetopause and around 
the edge of a plasmoid. The generation of these structures is related to the violation of Gauss's law. 
ECSIM-3 tries to satisfy Gauss's law better by incorporating the pseudo-current term, however, it 
creates oscillations in the magnetosphere just as iPIC3D does. ECSIM-2 produces smoother electric field profile in the magnetosheath than ECSIM-1. To demonstrate the role of $\theta$, we performed two more simulations 
for ECSIM-1 and ECSIM-2 with $\Delta t = 0.1\,s$ (corresponding to CFL$\approx 0.5$) instead of $\Delta t = 0.05\,s$ and show the results in 
Figure~\ref{fig:mp_ex_th}. ECSIM-1 generates wave-like structures in the magnetosheath, while the result 
of ECSIM-2 is still clean. Because of the significant improvement from $\theta=0.5$ to $\theta=0.51$, we 
use $\theta=0.51$ as our default value in practice. 

\begin{figure}
\begin{center}
\includegraphics[width=\textwidth]{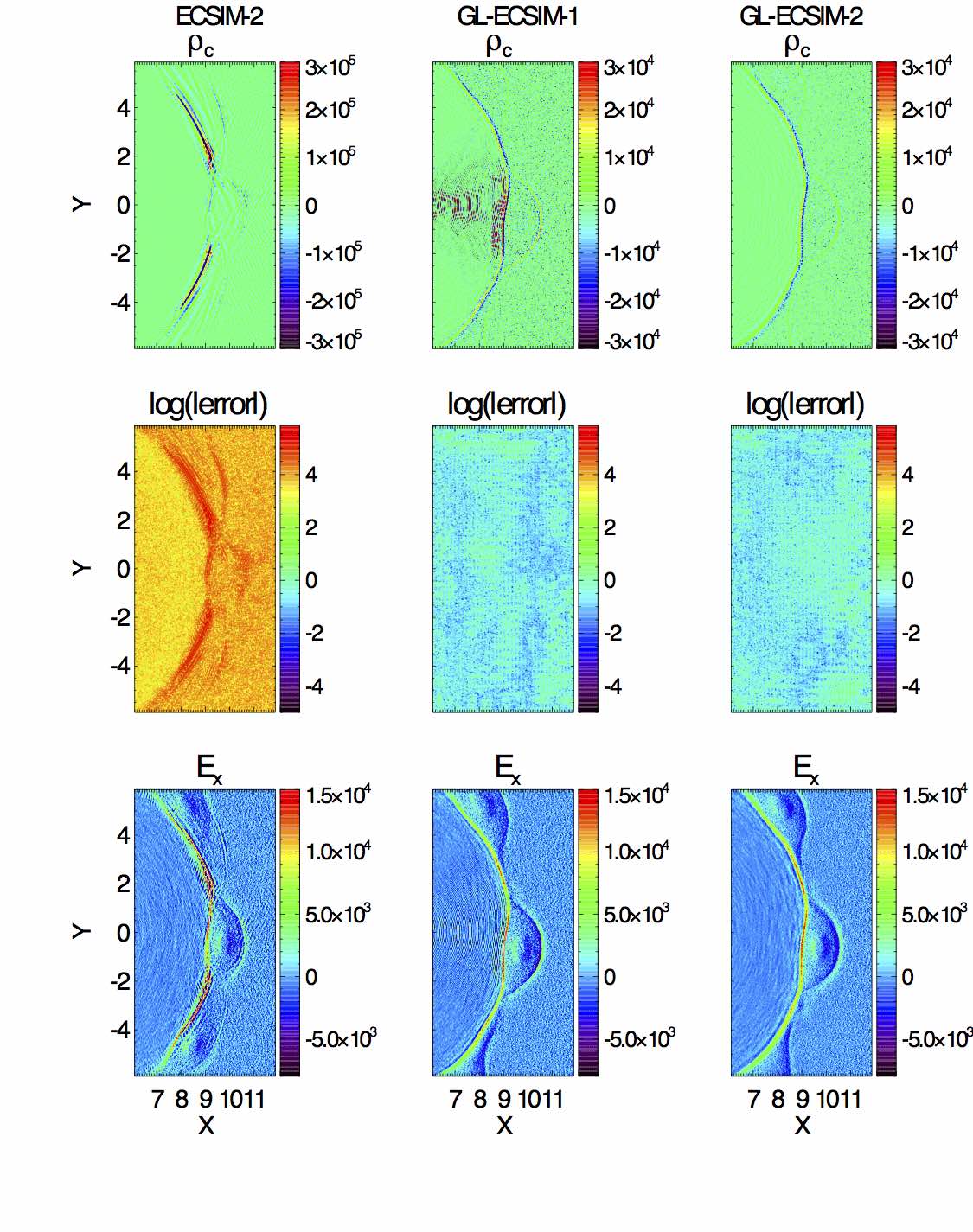}

\caption{The 2D magnetopause simulations with different parameters. From top to bottom: the net charge $\rho_c[\text{nT}/s]$, the absolute value of the error in Gauss's law, defined as $\nabla \cdot \mathbf{E}^{n+1}/(4\pi) - \rho_c^{n+1}$ with units $[\text{nT}/s]$, in logarithmic scale, and 
the electric field $E_x[\text{nT}\cdot\text{km/s}]$. 
From left to right: ECSIM with $\theta=0.51$, GL-ECSIM using compact discretization only, and GL-ECSIM with
extended stencil for the $\nabla \cdot \mathbf{E}$ discretization. See Table \ref{tb:parameters} for more 
details about the parameters.
We note that the color bar scale of the net charge density $\rho_c$ for ECSIM-2 is different from that of the others.  
\label{fig:mp_dive1}
}
\end{center}
\end{figure}

\begin{figure}
\begin{center}
\includegraphics[width=\textwidth]{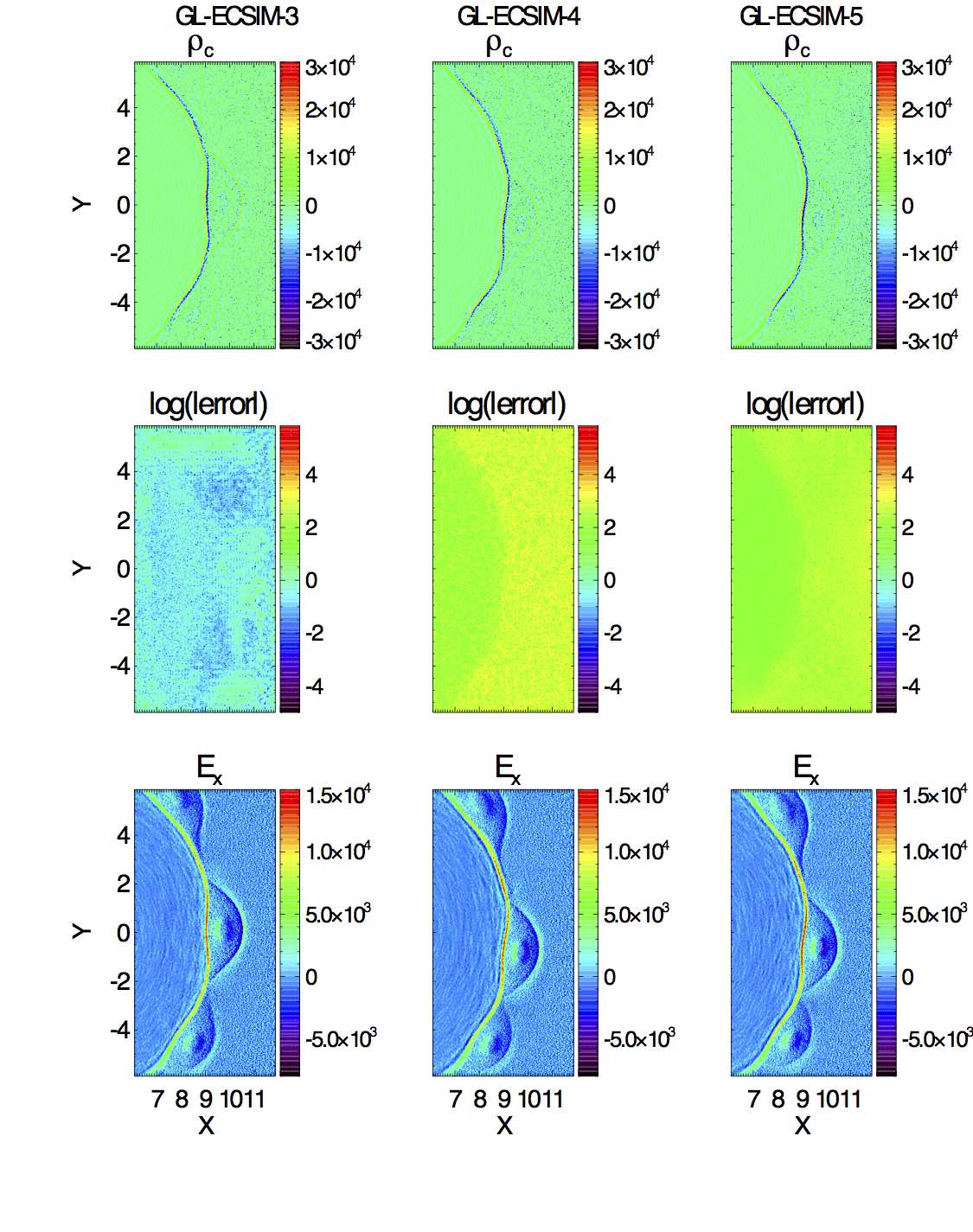}

\caption{The same variables as in Figure \ref{fig:mp_dive1}. From left to right: the accurate correction
for all species, the approximate global correction, and the approximate local correction. 
\label{fig:mp_dive2}
}
\end{center}
\end{figure}

Figure~\ref{fig:mp_dive1} and Figure~\ref{fig:mp_dive2} show the importance of satisfying Gauss's law and 
compare different particle position correction methods. We define the error in Gauss's law 
as $\nabla \cdot \mathbf{E}^{n+1}/(4\pi) - \rho_c^{n+1}$. For ECSIM-2, the net charge density and the error 
are the same order, which suggests Gauss's law is already dramatically violated. After the accurate correction 
method is applied to electrons to fix the Gauss's law error (GL-ECSIM-1), the error reduces to about 
$10^{-1}\, \text{[nT/s]}$, which is about 5 orders smaller than the net charge density. GL-ECSIM-1 also 
eliminates most of the small scale structures in ECSIM-2, such as the $E_x$ oscillations near the
edge of a plasmoid in the magnetosheath, but GL-ECSIM-1 produces significant short-wavelength oscillations
near the magnetopause at the magnetosphere side in the $E_x$ and net charge density $\rho_c$ profiles.
By using the extended stencil $\nabla \cdot \mathbf{E}$ spatial discretization (GL-ECSIM-2), these spurious 
oscillations are suppressed. Applying the accurate correction to all the species (GL-ECSIM-3) also 
obtain small error and smooth solution. The approximate global correction (GL-ECSIM-4) and 
approximate local correction (GL-ECSIM-5) can not perfectly eliminate the error in Gauss's law, 
and the errors are about $10^3\,\text{[nT/s]}$, which is 10 times smaller than the net charge $\rho_c$. 
Although the errors in GL-ECSIM-4 and GL-ECSIM-5 with the approximate correction are much larger than the errors in 
GL-ECSIM-2 and GL-ECSIM-3 employing the accurate correction, these four simulations produce results of similar quality. 

For this 2D magnetosphere test with the numerical parameters described above, the typical maximum particle 
displacement for GL-ECSIM-2 that corrects the electron particle positions only is about $4.0\%$ of 
the cell size after the first linear solve, $0.2\%$ after the second, and $0.03\%$ after the final third solve 
which is the end of the non-linear correction. When both electron and proton particle positions are 
corrected (GL-ECSIM-3), the corrections are half of these values. For the approximate correction methods 
(GL-ECSIM-4 and GL-ECSIM-5) the typical correction is about $2\%$ of the cell size.

These tests demonstrate GL-ECSIM method is more robust and accurate than iPIC3D and also the original 
ECSIM for a challenging problem. Since the pseudo-current method does not work well in general, $\theta=0.51$ is
more robust than $\theta=0.5$, and the extended stencil discretization of the $\nabla\cdot\mathbf{E}$ helps to 
suppress spurious oscillations, we will ignore the pseudo-current method, use $\theta=0.51$ and the extended stencil 
discretization with $c_{cpt}=0.9$ as default in the following tests. 

\subsection{Two-dimensional double-current-sheet magnetic reconnection}
The two-dimensional magnetic reconnection problem is widely used to test plasma simulation codes. 
The double-current-sheet setup allows periodic boundary conditions for both directions. Here we use
a setup based on the GEM-challenge \cite{Birn:2001}. 

The initial condition is set to satisfy the fluid force balance for both electrons and ions \cite{Huang:2018}.
The simulation domain is $-12.8 < x < 12.8$ and $-6.4 < y < 6.4$ in normalized CGS unit. The speed of light is
set to be $c=1$. The ion density is uniform and $n_i = 0.0975$. The ion plasma frequency is 
$\omega_{pi} = \sqrt{\frac{4 \pi n_i e^2}{m_i}} = 1.107$ and the ion inertial length 
$d_i = c/\omega_{pi} = 0.903$ since $m_i=1$ and $q_i=-q_e=1$. A reduced ion-electron mass ratio $m_i/m_e=25$ is used, 
so the electron skin depth is about $d_e=d_i/5=0.18$. Initially, there is no charge 
separation, $n_e=n_i$, and the electric field is $\mathbf{E} = 0$. 

The background magnetic filed is
 \begin{linenomath}
 \begin{eqnarray}
 B_x = B_0\left(-1 + \tanh\frac{y-y_B}{\delta} +
 \tanh\frac{y_T-y}{\delta}\right)
\label{eq:mr-bx}
 \end{eqnarray}
\end{linenomath}
where $B_0 = 0.07$, the positions of the two current sheets are $y_B=-3.2$ and $y_T = 3.2$, respectively, and the 
width of the currents sheets are controlled by $\delta = 0.5$. The electrons have a velocity in the z-direction
to generate current equal to the curl of the magnetic field, i.e, 
$J_z = n_e q_e u_{e,z} = -\partial B_x/\partial y$. The ion pressure $p_i$ is uniform in the whole domain.
Far away from the current sheets, the ion plasma beta is 1, and the electron pressure is 1/5 of the ion pressure. 
Near the current sheet, the electrons are heated to balance the magnetic field gradient force, which is 
the same as the Lorentz force $-n_e q_e u_{e,z}B_x$. This unperturbed initial condition is in fluid 
force balance \cite{Huang:2018}. 

A perturbation is added to excite the reconnection \cite{Lapenta:2017c}. The magnetic field perturbation vector
potential is $A_x = 0$, $A_y = 0$ and:
 \begin{linenomath}
 \begin{equation}
 \begin{split}
A_z = A_0 B_0 \bigg\lbrace 
 & -e^{-\frac{(x-x_T)^2}{G_x^2} 
       -\frac{(y-y_T)^2}{G_y^2}}
   \,\cos\left[k_x(x-x_T)\right]
     \cos\left[k_y(y-y_T)\right] \\
 & +e^{-\frac{(x-x_B)^2}{G_x^2}
       -\frac{(y-y_B)^2}{G_y^2}}
   \,\cos\left[k_x(x-x_B)\right]
     \cos\left[k_y(y-y_B)\right] 
\bigg\rbrace
\label{eq:mr-az}
\end{split}
 \end{equation}
\end{linenomath}
where the perturbation amplitude is set by $A_0 = 0.1$, the locations along the top and bottom current sheets are $x_T=6.4$ and $x_B = -6.4$, respectively, the width of Gaussian profiles are $G_x = G_y = 0.5$, and the wave vectors are $k_x = 2\pi/25.6$ and 
$k_y = 2\pi/12.8$. Since these two reconnection sites, i.e., the bottom left one at $(x_B, y_B)$ and the
top right one at $(x_T, y_T)$, produce the same signatures, we
only plot and discuss the bottom left reconnection site for simplicity.  

\begin{figure}
\begin{center}
\includegraphics[width=1.3\textwidth, trim=4cm 0cm 0cm 0cm]{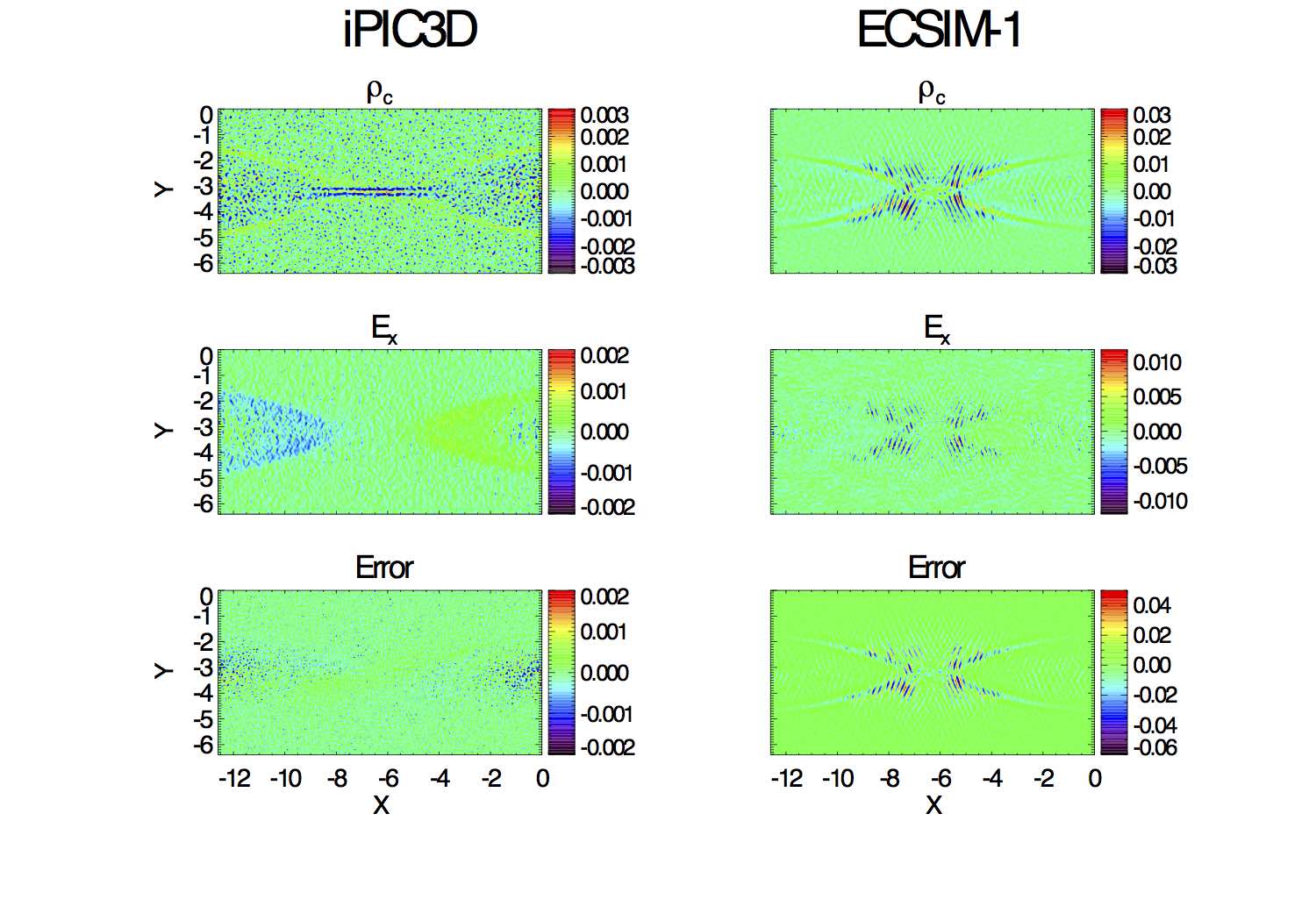}

\caption{The bottom left reconnection site of the double-current-sheets reconnection simulations at $t = 400$. The left panels show the iPIC3D simulation results, and the right panels show the results of ECSIM with $\theta=0.5$. From top to bottom: the net charge $q$, the electric field $E_x$ and the error in Gauss's law, 
defined as $\nabla \cdot \mathbf{E}^{n+1}/(4\pi) - \rho_c^{n+1}$. 
All these variables are in normalized units. $q$ and $E_x$ have the same units. The cell size is 
$\Delta x = 0.05$, and the time step is $\Delta t = 0.1 $. The results of ECSIM with 
$\theta = 0.51$ are not presented here, but they are very similar to the right panels above. 
\label{fig:mr_ipic3d_vs_ecsim}
}
\end{center}
\end{figure}

\begin{figure}
\begin{center}
\includegraphics[width=1.1\textwidth, trim=1cm 0cm 0cm 0cm]{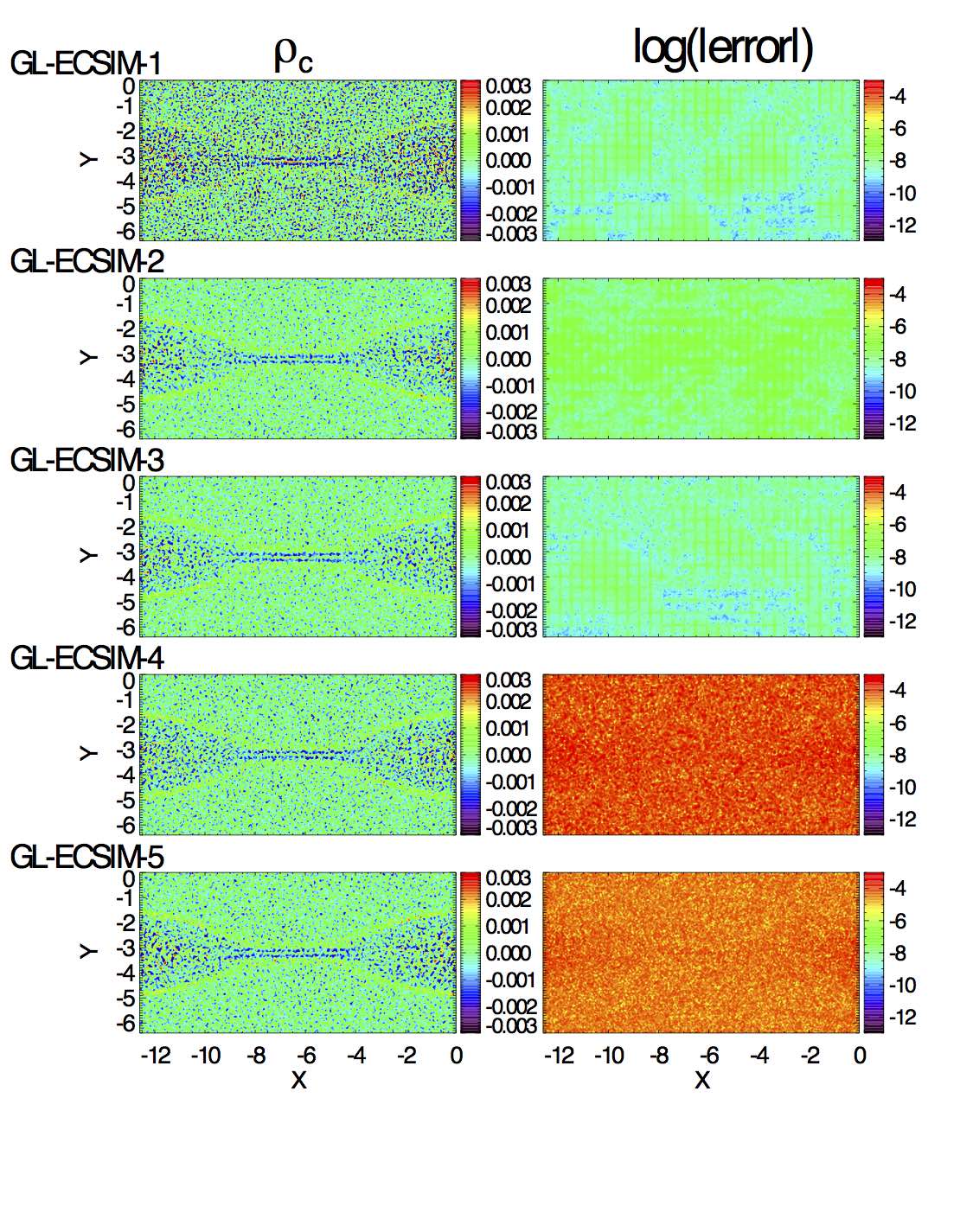}

\caption{The net charge density $q$ (left panel) and the absolute value of the error 
$\nabla \cdot \mathbf{E}^{n+1}/(4\pi) - \rho_c^{n+1}$ in Gauss's law (right panel) in logarithmic scale. 
The results for different parameters (see Table \ref{tb:parameters}) are presented from top to bottom.
\label{fig:mr_dive}
}
\end{center}
\end{figure}

\begin{figure}
\begin{center}
\includegraphics[width=\textwidth]{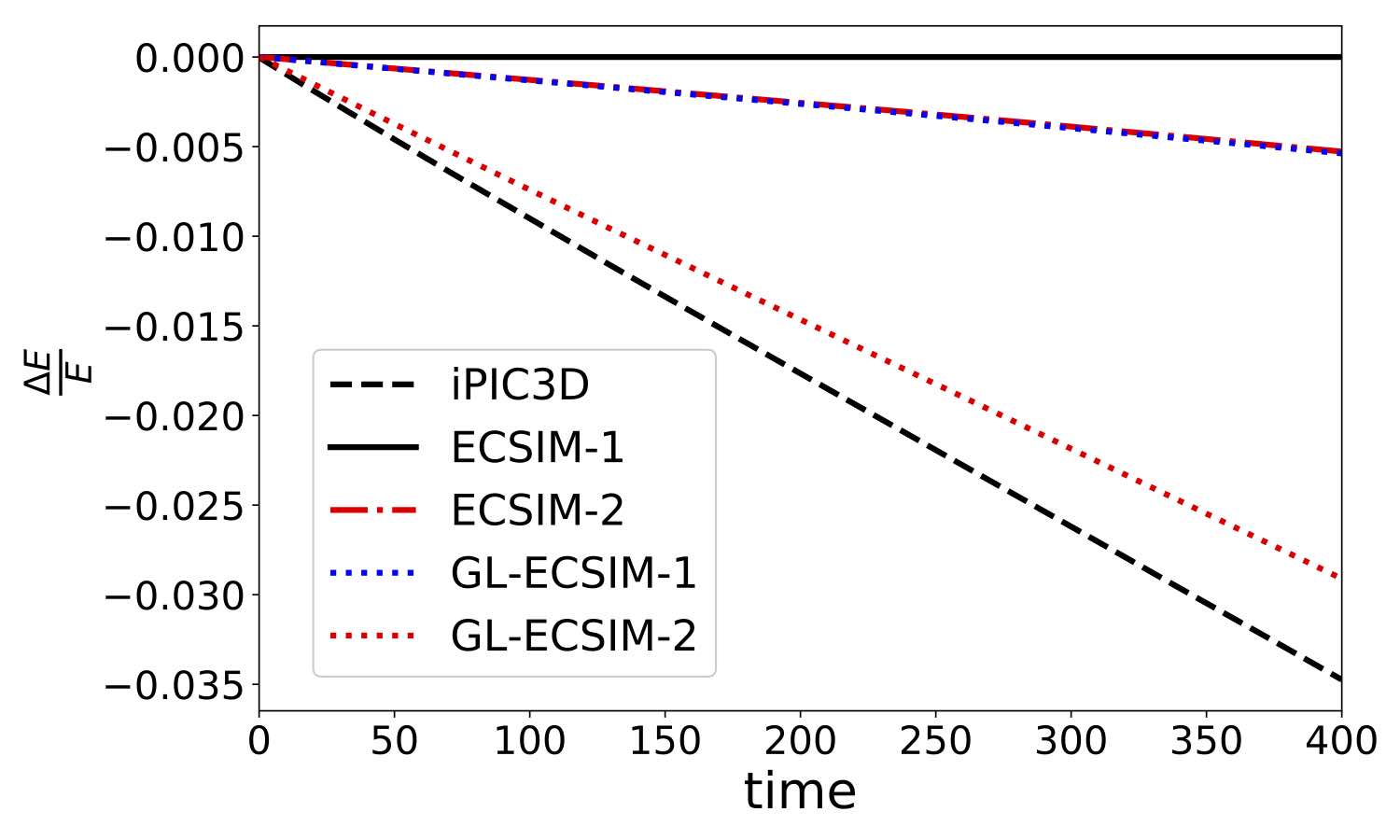}

\caption{The total energy variation of the 
double-current-sheet simulations for 
different schemes (see Table \ref{tb:parameters}). 
\label{fig:dE}
}
\end{center}
\end{figure}

For the simulations shown in Figures~\ref{fig:mr_ipic3d_vs_ecsim}, \ref{fig:mr_dive} and \ref{fig:dE}, 
the grid resolution is $\Delta x = 0.05$ and the time step is $\Delta t = 0.1$. There are 900 macro-particles per cell per species.
 The simulation results at $t=400$ are shown. Figure~\ref{fig:mr_ipic3d_vs_ecsim} shows 
 the net charge $\rho_c$, electric field $E_x$ and the error in Gauss's law 
for iPIC3D and ECSIM. iPIC3D produces good quality results for this test. Near the reconnection 
site, the divergent field-aligned electric field $E_x$ is well resolved, a double-sandwich
structure of the net charge in the center of the reconnection site is captured, and the 
error is small and dominated by the random particle noise.
However, $\rho_c$ and $E_x$ of ECSIM are dominated by the unphysical
 oscillations along the separatrices, and the huge error indicates that Gauss's law is dramatically violated. 
The ECSIM simulation shown here uses $\theta=0.5$, and the simulation with
$\theta=0.51$ does not alleviate the issue. To the best of our knowledge, the double-sandwich net charge structure has not been discussed 
in detail before, but some published high-resolution PIC simulations have provided the 
evidence of the existence of this structure. For example, the static electric field 
$E_z$, which is $E_y$ in the current paper, shown in 
Figure~1(a) of \cite{Guo:2017} and Figure~3(b) of \cite{Ng:2012} 
is corresponding to the double-sandwich net charge.

Comparing the GL-ECSIM-1 and GL-ECSIM-2 results in Figure~\ref{fig:mr_dive} demonstrates that the extended stencil discretization of 
$\nabla \cdot \mathbf{E}$ helps
to reduce the noise. All the position correction methods produce essentially the same net charge structure (GL-ECSIM-2 to GL-ECSIM-5). The error in Gauss's law is about 5 orders smaller than the net charge density in the simulations employing
the accurate correction method (GL-ECSIM-1 to GL-ECSIM-3), and it is about 1 order smaller for the approximate corrections 
(GL-ECSIM-4 and GL-ECSIM-5). 
When the accurate correction is only applied to electrons (GL-ECSIM-1 and GL-ECSIM-2), the typical
maximum particle displacement is $4.5\%$, $0.12\%$ and $0.002\%$ of the cell size for the three linearized corrections. These values reduce by a factor of 2 when both electrons
 and ions are corrected (GL-ECSIM-3). 
 For the approximate corrections 
 GL-ECSIM-4 and GL-ECSIM-5, the
 typical maximum displacement is about $3\%$ of the cell size.   
 
 Figure~\ref{fig:dE} shows the total energy variation. For ECSIM with $\theta=0.5$ (ECSIM-1), the energy is conserved, the small error corresponds to the accuracy of the iterative implicit electric field solver. ECSIM with $\theta=0.51$ (ECSIM-2) 
 dissipates $0.5\%$ of the total energy after 4000 iterations. The plots of GL-ECSIM-1 and ECSIM-2 
 are overlapped with each other because the particle position correction does not change the energy. The extended 
 stencil discretization of $\nabla \cdot \mathbf{E}$ (GL-ECSIM-2) dissipates $3\%$ of the energy, which is still 
 a relatively small value. The energy variation for other correction methods (GL-ECSIM-3 to GL-ECSIM-5) 
 are essentially the same as GL-ECSIM-2. As a comparison, the total energy of the iPIC3D simulation reduces about $3.5\%$
 
  The normalized wall time for each simulation is presented in Table~\ref{tb:parameters}. 
  From the timing results, we conclude:
\begin{itemize}
\item  In our implementation, ECSIM is about twice slower than iPIC3D.
\item  For the accurate correction method (GL-ECSIM-1 to GL-ECSIM-3), the correction 
takes $30\%$ to $40\%$ of the total simulation time.
\item  Correcting all species (GL-ECSIM-3) is about $10\%$ slower than correcting one species 
only (GL-ECSIM-1 and GL-ECSIM-2). 
\item The approximate correction methods only take about $10\%$ or less of the total wall time. 
\end{itemize}  
In practice, we prefer the approximate global correction method since it reaches a balance between
robustness and efficiency. The approximate local correction method is even faster, but it is less robust and accurate for some challenging problems.   
 
\begin{figure}
\begin{center}
\includegraphics[width=1.2\textwidth, trim=3cm 0cm 0cm 0cm ]{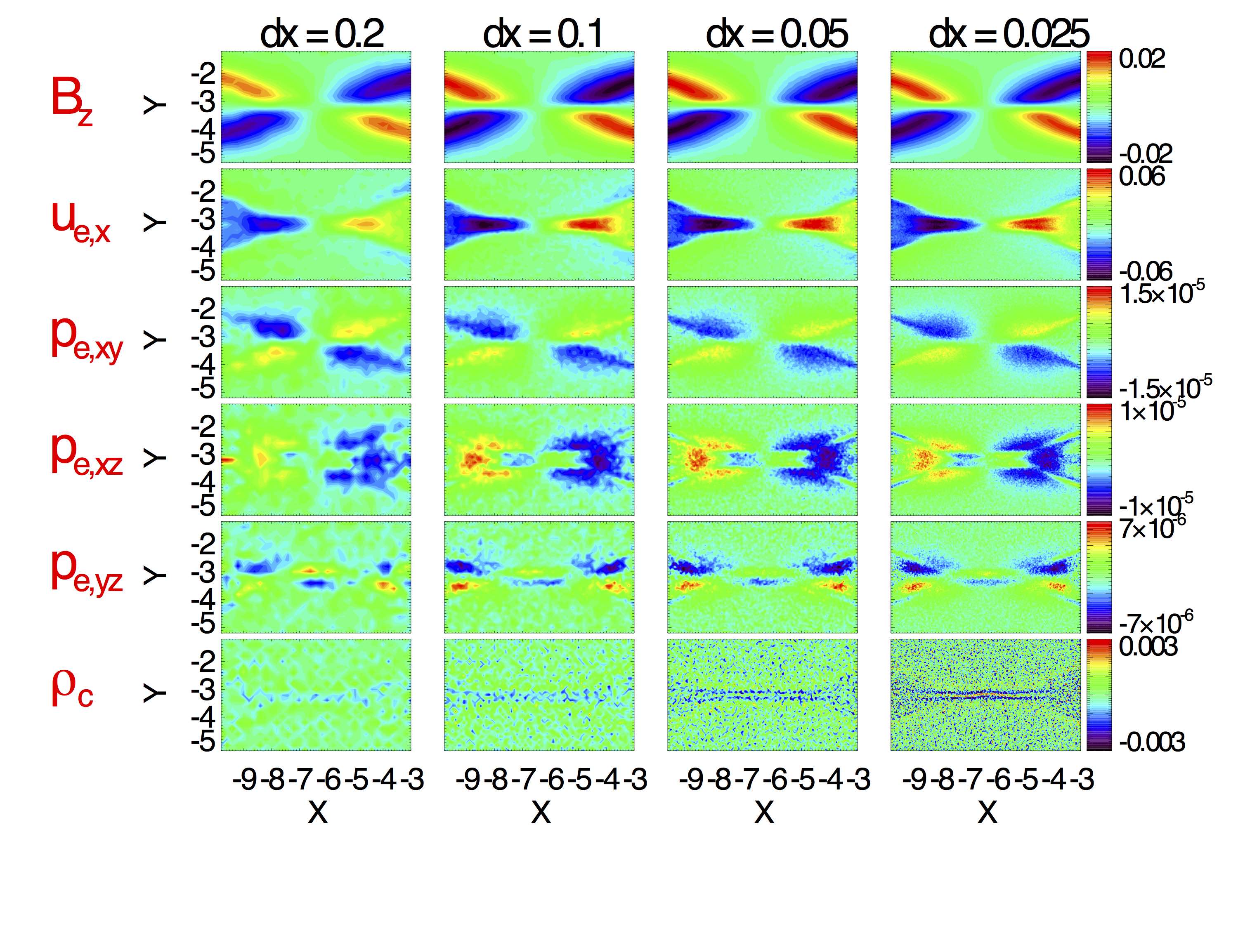}

\caption{The grid convergence study of the 
double-current-sheet simulation with the approximate global 
correction method (GL-ECSIM-4 in Table \ref{tb:parameters}). From top to bottom: the out-of-plane 
Hall magnetic field $B_z$, the electron jet velocity $u_{e,x}$, the three electron off-diagonal 
pressure tensor elements $p_{e,xy}$, $p_{e,xz}$ and $p_{e,yz}$, and the net charge density $\rho_c$ at $t=400$ are 
shown in normalized units. From left to right, the cell sizes are $\Delta x = 0.2$, $\Delta x = 0.1$, 
$\Delta x = 0.05$ and $\Delta x = 0.025$, and the corresponding time steps are $\Delta t = 0.4$, 
$\Delta t = 0.2$, $\Delta t = 0.1$ and $\Delta t = 0.05$, respectively.
\label{fig:mr_converge}
}
\end{center}
\end{figure}

\begin{figure}
\begin{center}
\includegraphics[width=\textwidth]{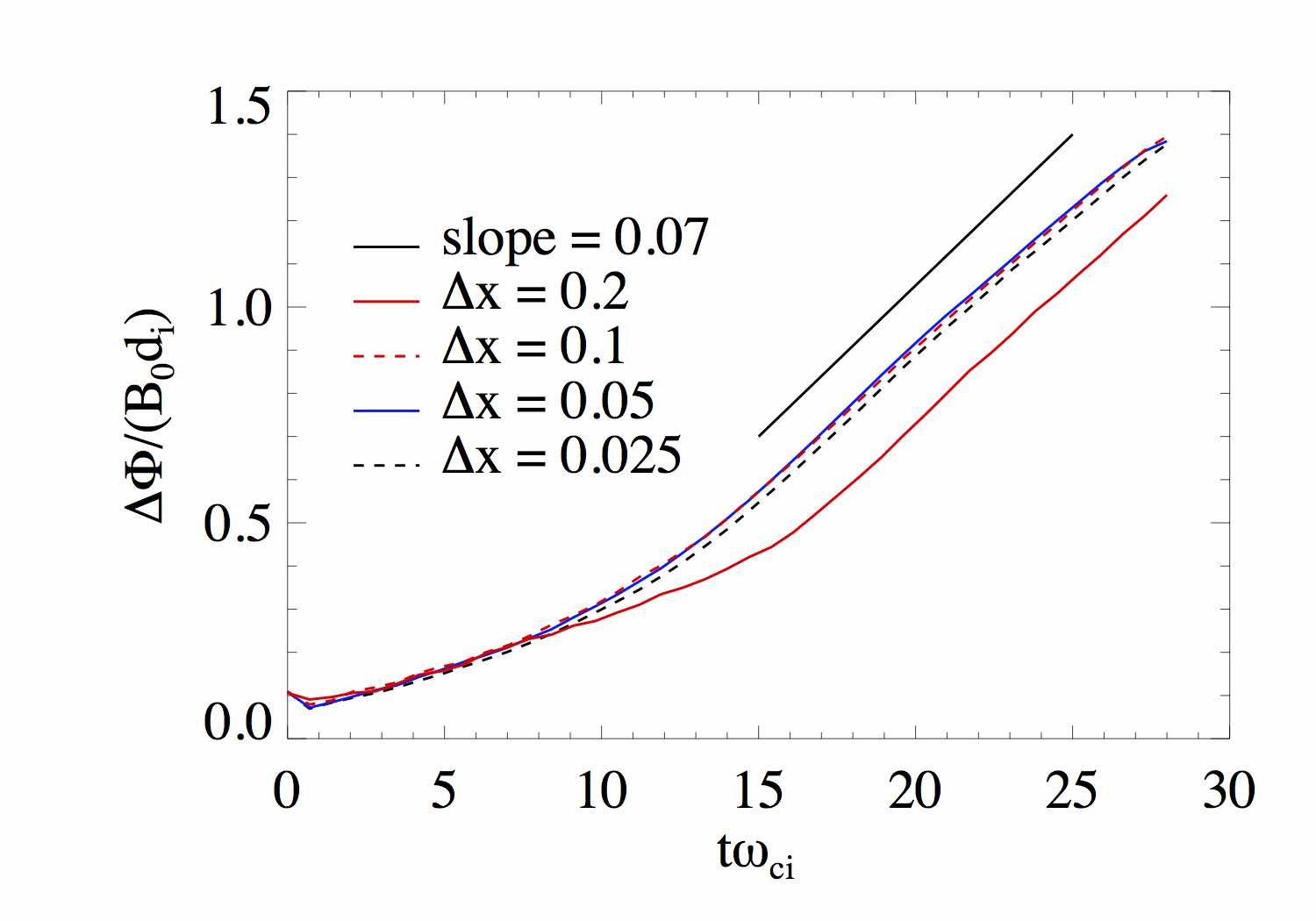}

\caption{The reconnection rate for the simulations shown in Figure \ref{fig:mr_converge}. All simulations have a 
reconnection rate of $\sim 0.07$. 
\label{fig:mr_rate}
}
\end{center}
\end{figure}

Figure~\ref{fig:mr_converge} shows the results with the approximate global correction for
grid resolution 0.2, 0.1, 0.05 and 0.025. The CFL number is fixed and the corresponding time 
steps are 0.4, 0.2, 0.1 and 0.05, respectively. All the simulations capture the Hall magnetic field $B_z$, even the electron
flows, such as $u_{e,x}$, very well. Once the grid resolution is close to or higher than half of the electron 
skin depth $d_e=0.18$, the details of the off-diagonal electron pressure terms are also well resolved,
while the simulation with $\Delta x = 0.2$ is too diffusive to capture these details. 
The pressure components presented
here is similar to other high-resolution PIC simulations, such as the Figure~9 in \cite{Wang:2015}. 
The double-sandwich structure of the net charge is even harder to capture. Even the simulation with
$\Delta x = 0.1$ does not resolve this structure well. The normalized reconnection rate is shown
in Figure~\ref{fig:mr_rate}. The four simulations with different grid resolution have the same 
normalized reconnection rate of 0.07. The algorithm to calculate the reconnection rate can be 
found in \cite{Huang:2018}. These four simulations demonstrate that the GL-ECSIM method converges well with increasing grid resolution, and a variety of
reconnection related structures can be captured once the grid resolution is close to or higher 
than half of the electron skin depth. 

\subsection{Weibel instability}
Finally, we perform the 1D Weibel instability test to quantitatively prove that the 
particle correction methods do not interfere
with properly capturing the growth and evolution of this instability.

\begin{figure}
\begin{center}
\includegraphics[width=\textwidth]{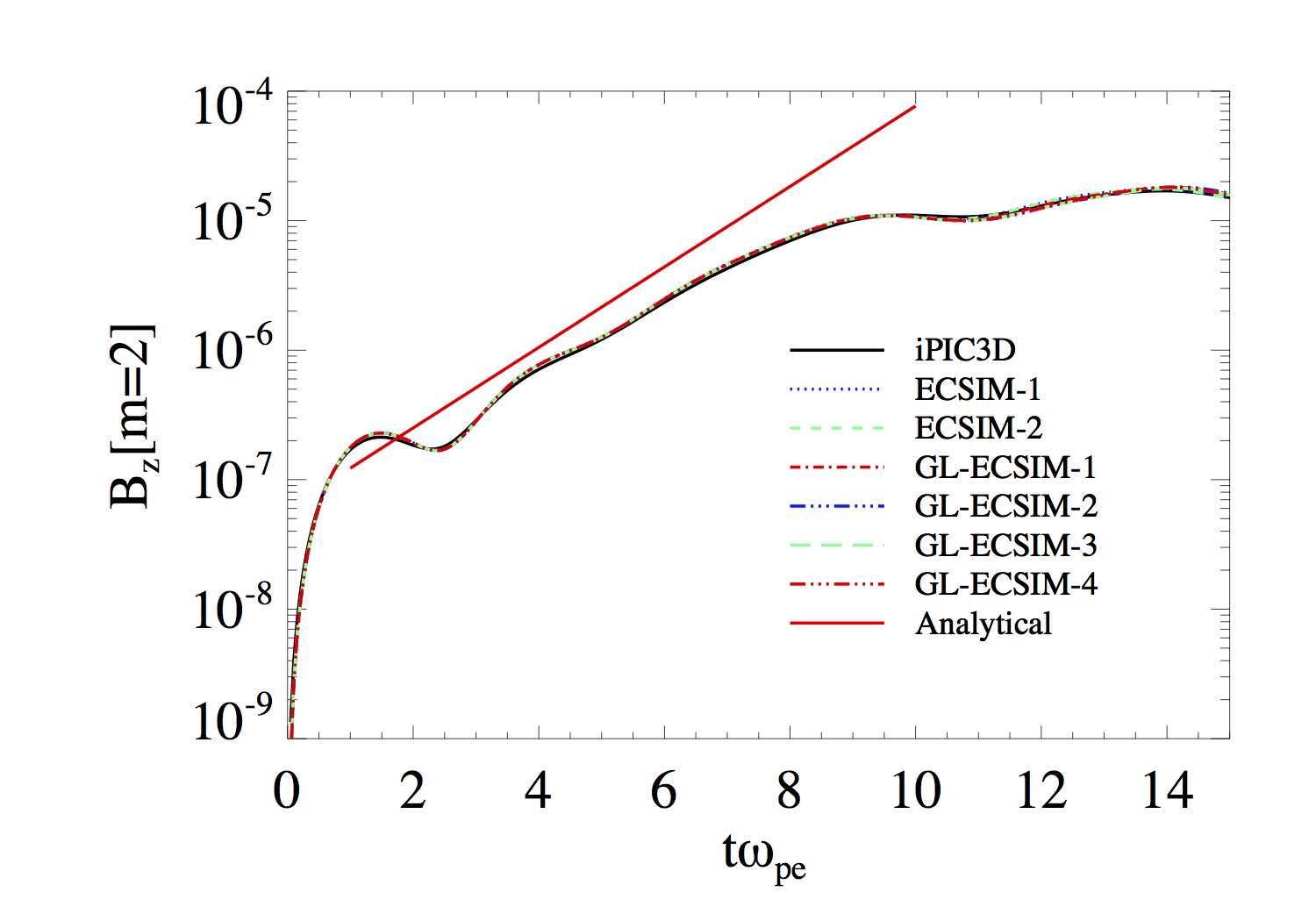}

\caption{The growth of the Weibel instability. The analytic growth rate is $\gamma = 0.716\omega_{pe}$. 
The particle correction methods do not change the growth rate at all. 
\label{fig:weilel}
}
\end{center}
\end{figure}

The simulation is performed on a 1D domain of size $L_x = 2\pi d_e$, resolved by cells of size 
$\Delta x = L_x/64$ and time step $\Delta t = 0.05/\omega_{pe}$. 400 particles per cell per species 
are used. Each of the two counter-streaming electron beams has a speed of $0.8c$ along 
the positive or negative y-direction. The thermal velocity of the electrons is $u_{e,th} = 0.01c$. 
The ions are uniformly distributed to satisfy the charge neutrality requirement, but the ions 
are much colder and heavier than the electrons ($m_i/m_e=10^4$ and $u_{i,th} = 10^{-8}c$), so that 
the ions do not move essentially. The linear theory \cite{Weibel:1959} predicts the growth 
rate of the mode with wavelength $\pi d_e$ is $\gamma = 0.716\omega_{pe}$. Figure~\ref{fig:weilel}
shows that the growth rates are essentially the same for all the simulations, and the rate is close to the analytic value during
the linear growth stage. 

\section{Conclusion}
\label{section:conclusion}
In this paper, we introduce the novel GL-ECSIM algorithm, which can satisfy both the total 
energy conservation and 
Gauss's law to the accuracy of the iterative solvers. In practice, we need to sacrifice the energy
conservation a little bit and introduce a small amount of  
diffusion to reduce noise and suppress numerical oscillations by using a time centering parameter $\theta=0.51$ instead of 0.5 of the original ECSIM algorithm. In addition, we introduce a linear combination of the original compact 
stencil (with a 0.9 weight) and a new extended stencil (with 0.1 weight) for the discretization of the $\nabla \cdot \mathbf{E}$ term in the electric field equation. In effect, this adds a dissipation term proportional to the 4th derivative of the electric field, which helps to remove spurious oscillations.

Our 2D reconnection and magnetosphere simulations suggest that the original ECSIM scheme may produce numerical 
artifacts due to the violation of Gauss's law. In order to solve this problem without changing
the energy, we design a class of new algorithms to correct the particle positions after each ECSIM update to satisfy 
Gauss's law. The accurate correction method carefully calculates the displacement of each 
particle to eliminate the error in Gauss's law accurately while minimizing the norm of the total displacements. This accurate correction method requires a non-linear iterative solver and 
takes $30\%$ to $40\%$ of the total wall time to do the correction. In order to speed up the simulation,
we introduce another two approximate methods. The approximate global correction method solves a Poisson's 
equation to estimate the particle displacement, and the approximate local correction estimates 
the displacement based on the surrounding errors. The local correction method is faster than 
the global correction. But the global correction calculate the displacement based on global 
information, which makes the global correction more robust for challenging problems.

Using the approximate global GL-ECSIM method with its optimal parameter settings, we performed a grid convergence study for the magnetic reconnection problem. We found that the solution converges well with diminishing grid resolution, and it is converged in most variables if the grid resolution is about one half of the electron skin depth.

Our tests demonstrate that the GL-ECSIM is robust and accurate. It has been successfully applied to our ongoing 3D global magnetospheric simulations.

\vspace{5mm}
{\noindent \bf Acknowledgments:}
This work was supported by the INSPIRE NSF grant PHY-1513379 and the NSF
PREEVENTS grant 1663800. Computational resources supporting this work were 
provided on the Blue Waters super computer by the NSF PRAC grant ACI-1640510,
on the Pleiades computer by NASA High-End Computing (HEC) Program through 
the NASA Advanced Supercomputing (NAS) Division at Ames Research Center,
and from Yellowstone (ark:/85065/d7wd3xhc) provided by NCAR's Computational and
Information Systems Laboratory, sponsored by the National Science Foundation.

\newpage
\bibliographystyle{elsarticle-num}
\bibliography{csem,yuxichen}






\end{document}